%% file: main.tex
\documentclass[12pt,a4paper]{article}

\usepackage{ifthen} 
\newboolean{pdflatex}
\setboolean{pdflatex}{true} 

\newboolean{articletitles}
\setboolean{articletitles}{true} 

\newboolean{uprightparticles}
\setboolean{uprightparticles}{false} 

\newboolean{inbibliography}
\setboolean{inbibliography}{false} 

\input{preamble}
\usepackage{longtable} 

\begin{document}

\renewcommand{\thefootnote}{\fnsymbol{footnote}}
\setcounter{footnote}{1}

\input{title-LHCb-PAPER}


\renewcommand{\thefootnote}{\arabic{footnote}}
\setcounter{footnote}{0}



\pagestyle{plain} 
\setcounter{page}{1}
\pagenumbering{arabic}


%

\input{introduction}

\input{setup}

\input{measurements}
\input{results}
\input{conclusions} 
\input{acknowledgements}
\input{references}






\end{document}

%% file: preamble.tex
\usepackage{microtype}
\usepackage{lineno}  
\usepackage{xspace} 
\usepackage{caption} 

\usepackage{graphicx}  
\usepackage{color}
\usepackage{colortbl}
\graphicspath{{./figs/}} 

\usepackage{amsmath} 
\usepackage{amssymb}
\usepackage{amsfonts}
\usepackage{upgreek} 

\newcommand*\patchAmsMathEnvironmentForLineno[1]{%
\expandafter\let\csname old#1\expandafter\endcsname\csname #1\endcsname
\expandafter\let\csname oldend#1\expandafter\endcsname\csname
end#1\endcsname
 \renewenvironment{#1}%
   {\linenomath\csname old#1\endcsname}%
   {\csname oldend#1\endcsname\endlinenomath}%
}
\newcommand*\patchBothAmsMathEnvironmentsForLineno[1]{%
  \patchAmsMathEnvironmentForLineno{#1}%
  \patchAmsMathEnvironmentForLineno{#1*}%
}
\AtBeginDocument{%
\patchBothAmsMathEnvironmentsForLineno{equation}%
\patchBothAmsMathEnvironmentsForLineno{align}%
\patchBothAmsMathEnvironmentsForLineno{flalign}%
\patchBothAmsMathEnvironmentsForLineno{alignat}%
\patchBothAmsMathEnvironmentsForLineno{gather}%
\patchBothAmsMathEnvironmentsForLineno{multline}%
\patchBothAmsMathEnvironmentsForLineno{eqnarray}%
}

\usepackage{hyperref}    
\usepackage[all]{hypcap} 

\input{lhcb-symbols-def} 

\usepackage{cite} 
\usepackage{mciteplus}

%% file: lhcb-symbols-def.tex
\usepackage{xspace} 
\usepackage{upgreek}

\def\lhcb {\mbox{LHCb}\xspace}

\def\MagUp {\mbox{\em Mag\kern -0.05em Up}\xspace}

\ifthenelse{\boolean{uprightparticles}}%
{

 \def\PDelta      {\ensuremath{\Delta}\xspace}                 
 \def\PXi      {\ensuremath{\Xi}\xspace}                 
 \def\PLambda      {\ensuremath{\Lambda}\xspace}                 
 \def\PSigma      {\ensuremath{\Sigma}\xspace}                 
 \def\POmega      {\ensuremath{\Omega}\xspace}                 
 \def\PUpsilon      {\ensuremath{\Upsilon}\xspace}                 
 
 \def\PB      {\ensuremath{\mathrm{B}}\xspace}                 
                  
 \def\PD      {\ensuremath{\mathrm{D}}\xspace}

 \def\PK      {\ensuremath{\mathrm{K}}\xspace}

 \def\Pi      {\ensuremath{\mathrm{i}}\xspace}

}
{

 \mathchardef\PDelta="7101
 \mathchardef\PXi="7104
 \mathchardef\PLambda="7103
 \mathchardef\PSigma="7106
 \mathchardef\POmega="710A
 \mathchardef\PUpsilon="7107
                  
 \def\PB      {\ensuremath{B}\xspace}                 
                  
 \def\PD      {\ensuremath{D}\xspace}

 \def\PK      {\ensuremath{K}\xspace}

 \def\Pi      {\ensuremath{i}\xspace}

}

\makeatletter
\ifcase \@ptsize \relax
  \newcommand{\miniscule}{\@setfontsize\miniscule{4}{5}}
\or
  \newcommand{\miniscule}{\@setfontsize\miniscule{5}{6}}
\or
  \newcommand{\miniscule}{\@setfontsize\miniscule{5}{6}}
\fi
\makeatother

\DeclareRobustCommand{\optbar}[1]{\shortstack{{\miniscule (\rule[.5ex]{1.25em}{.18mm})}
  \\ [-.7ex] $#1$}}

  \def\Kbar    {{\kern 0.2em\overline{\kern -0.2em \PK}{}}\xspace}

\def\KorKbar    {\kern 0.18em\optbar{\kern -0.18em K}{}\xspace}

  \def\Dbar    {{\kern 0.2em\overline{\kern -0.2em \PD}{}}\xspace}

\def\DorDbar    {\kern 0.18em\optbar{\kern -0.18em D}{}\xspace}

\def\Bbar    {{\ensuremath{\kern 0.18em\overline{\kern -0.18em \PB}{}}}\xspace}

\def\BorBbar    {\kern 0.18em\optbar{\kern -0.18em B}{}\xspace}

  \def\Y#1S{\ensuremath{\PUpsilon{(#1S)}}\xspace}

\def\Lbar        {{\ensuremath{\kern 0.1em\overline{\kern -0.1em\PLambda}}}\xspace}
\def\LorLbar    {\kern 0.18em\optbar{\kern -0.18em \PLambda}{}\xspace}

\def\AT#1     {\ensuremath{A_{\mathrm{T}}^{#1}}\xspace}           

\def\C#1      {\ensuremath{\mathcal{C}_{#1}}\xspace}                       
\def\Cp#1     {\ensuremath{\mathcal{C}_{#1}^{'}}\xspace}                    
\def\Ceff#1   {\ensuremath{\mathcal{C}_{#1}^{\mathrm{(eff)}}}\xspace}        
\def\Cpeff#1  {\ensuremath{\mathcal{C}_{#1}^{'\mathrm{(eff)}}}\xspace}       
\def\Ope#1    {\ensuremath{\mathcal{O}_{#1}}\xspace}                       
\def\Opep#1   {\ensuremath{\mathcal{O}_{#1}^{'}}\xspace}                    

\newcommand{\jinst}[3]{\href{http://www.iop.org/EJ/abstract/1748-0221/#1/01/#3}{#2 {\it{JINST}} {\bf{#1}} #3}}

\newcommand{\tev}{\ifthenelse{\boolean{inbibliography}}{\ensuremath{~T\kern -0.05em eV}\xspace}{\ensuremath{\mathrm{\,Te\kern -0.1em V}}}\xspace}
\newcommand{\gev}{\ensuremath{\mathrm{\,Ge\kern -0.1em V}}\xspace}
\newcommand{\mev}{\ensuremath{\mathrm{\,Me\kern -0.1em V}}\xspace}
\newcommand{\kev}{\ensuremath{\mathrm{\,ke\kern -0.1em V}}\xspace}
\newcommand{\ev}{\ensuremath{\mathrm{\,e\kern -0.1em V}}\xspace}
\newcommand{\gevc}{\ensuremath{{\mathrm{\,Ge\kern -0.1em V\!/}c}}\xspace}
\newcommand{\mevc}{\ensuremath{{\mathrm{\,Me\kern -0.1em V\!/}c}}\xspace}
\newcommand{\gevcc}{\ensuremath{{\mathrm{\,Ge\kern -0.1em V\!/}c^2}}\xspace}
\newcommand{\gevgevcccc}{\ensuremath{{\mathrm{\,Ge\kern -0.1em V^2\!/}c^4}}\xspace}
\newcommand{\mevcc}{\ensuremath{{\mathrm{\,Me\kern -0.1em V\!/}c^2}}\xspace}

\def\textmu {\ensuremath{{\,\upmu}}\xspace}

\def\gsim{{~\raise.15em\hbox{$>$}\kern-.85em
          \lower.35em\hbox{$\sim$}~}\xspace}
\def\lsim{{~\raise.15em\hbox{$<$}\kern-.85em
          \lower.35em\hbox{$\sim$}~}\xspace}

\def\tell1  {TELL1\xspace}
\def\ukl1   {UKL1\xspace}



%% file: title-LHCb-PAPER.tex

\begin{titlepage}
\pagenumbering{roman}

\vspace*{-1.5cm}
\centerline{\large EUROPEAN ORGANIZATION FOR NUCLEAR RESEARCH (CERN)}
\vspace*{0.75cm}
\noindent
\begin{tabular*}{\linewidth}{lc@{\extracolsep{\fill}}r@{\extracolsep{0pt}}}
\ifthenelse{\boolean{pdflatex}}
{\vspace*{-2.7cm}\mbox{\!\!\!\includegraphics[width=.14\textwidth]{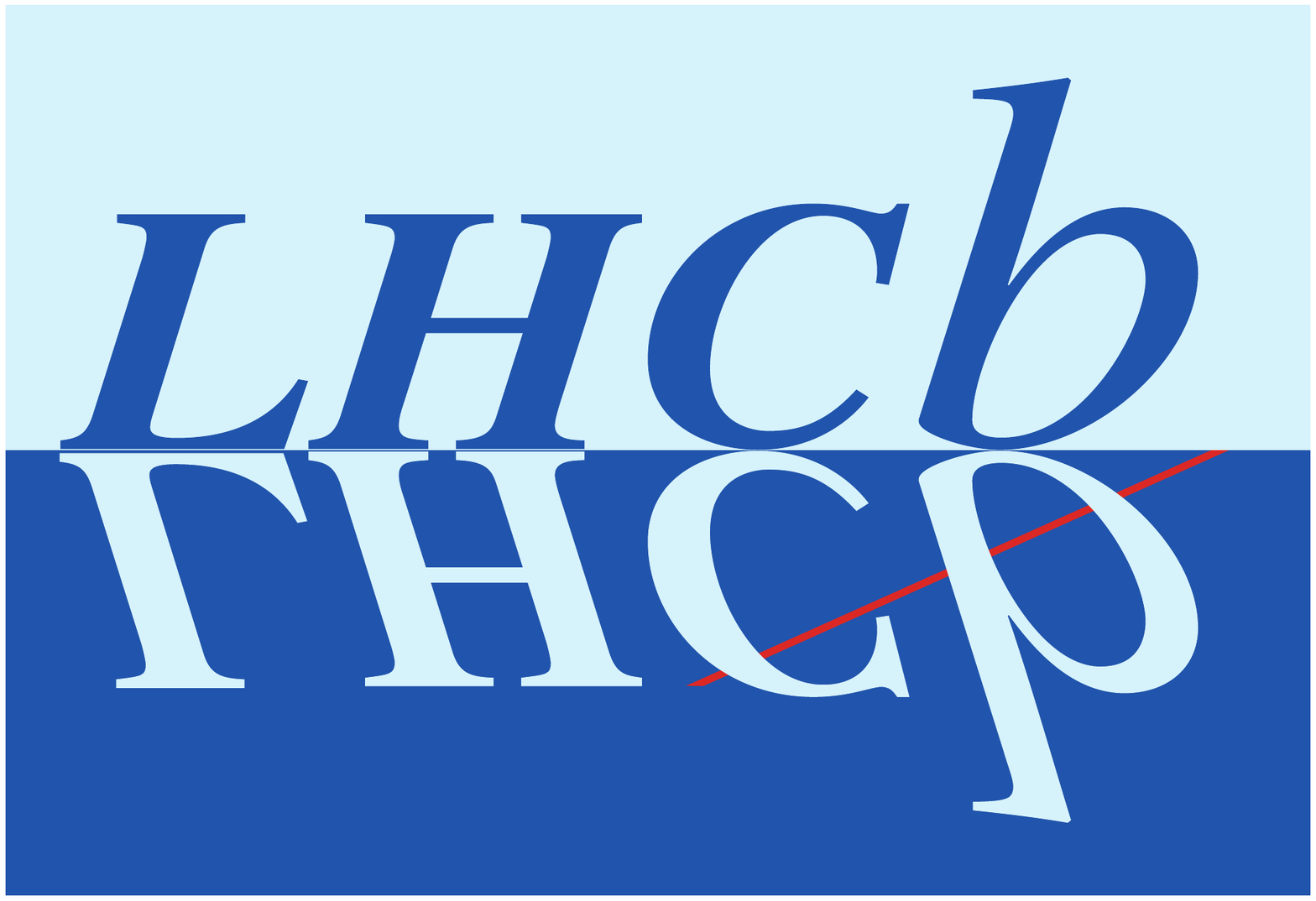}} & &}%
{\vspace*{-1.2cm}\mbox{\!\!\!\includegraphics[width=.12\textwidth]{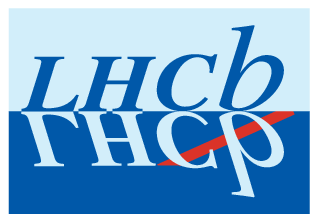}} & &}%
\\
 & & LHCb-PUB-2016-002 \\  
 & & \today \\ 
 & & \\
\end{tabular*}

\vspace*{0.25cm}

{\normalfont\bfseries\boldmath\huge
\begin{center}
Characterisation of Medipix3 \\Silicon Detectors in a Charged-Particle Beam
\end{center}
}

\vspace*{0.6cm}

\begin{center}
K.~Akiba$^1$, 
J.~Alozy$^2$, 
R.~Aoude$^3$, 
M.~van~Beuzekom$^4$, 
J.~Buytaert$^2$, 
P.~Collins$^2$, 
A.~Dosil~Su\'arez$^5$, 
R.~Dumps$^2$, 
A.~Gallas$^5$,
C.~Hombach$^6$,
D.~Hynds$^7$,
M.~John$^8$,
A.~Leflat$^9$,
Y.~Li$^{10}$,
E.~P\'erez~Trigo$^5$,
R.~Plackett$^{11}$,
M.~M.~Reid$^{12}$,
P.~Rodr\'iguez~P\'erez$^6$,
H.~Schindler$^2$\footnote{Corresponding author.},
P.~Tsopelas$^4$,
C.~V\'azquez~Sierra$^5$\footnote{Corresponding author.},
J.~J.~Velthuis$^{13}$, 
M.~Wysoki\'{n}ski$^{14}$
\bigskip\\
{\normalfont\itshape\footnotesize
$^1$ Universidade Federal do Rio de Janeiro, Rio de Janeiro, Brazil\\
$^2$ European Organization for Nuclear Research (CERN), Geneva, Switzerland\\
$^3$ Pontif\'icia Universidade Cat\'olica do Rio de Janeiro, Rio de Janeiro, Brazil\\
$^4$ Nikhef National Institute for Subatomic Physics, Amsterdam, The Netherlands\\
$^5$ Universidade de Santiago de Compostela, Santiago de Compostela, Spain\\
$^6$ School of Physics and Astronomy, University of Manchester, Manchester, United Kingdom\\
$^7$ School of Physics and Astronomy, University of Glasgow, Glasgow, United Kingdom\\
$^8$ Department of Physics, University of Oxford, Oxford, United Kingdom\\
$^9$ Institute of Nuclear Physics, Moscow State University (SINP MSU), Moscow, Russia\\
$^{10}$ Center for High Energy Physics, Tsinghua University, Beijing, China\\
$^{11}$ Diamond Light Source Ltd., Didcot, United Kingdom\\
$^{12}$ Department of Physics, University of Warwick, Coventry, United Kingdom\\
$^{13}$ H.H. Wills Physics Laboratory, University of Bristol, Bristol, United Kingdom\\
$^{14}$ AGH - University of Science and Technology, Faculty of Computer Science, Electronics and Telecommunications, Krak\'{o}w, Poland\\
}
\end{center}
\vspace{0.1cm}

\begin{abstract}
\noindent While designed primarily for X-ray imaging applications, the Medipix3 ASIC can also be used for charged-particle tracking. 
In this work, results from a beam test at the CERN SPS with irradiated and non-irradiated sensors are presented and shown to be in agreement with simulation, demonstrating the suitability of the Medipix3 ASIC as a tool for characterising pixel sensors.
\end{abstract}

\vspace*{0.75cm}

\begin{center}
  Published in JINST
\end{center}

\vspace{\fill}

{\footnotesize 
\centerline{\copyright~CERN on behalf of the \lhcb collaboration, licence \href{http://creativecommons.org/licenses/by/4.0/}{CC-BY-4.0}.}}
\vspace*{2mm}

\end{titlepage}


\newpage
\setcounter{page}{2}
\mbox{~}
%
%
%
%

\cleardoublepage

%% file: introduction.tex
\section{Introduction}
As part of the upgrade of the LHCb experiment, 
scheduled for the second long shutdown of the LHC in 2018/19,
the present microstrip-based Vertex Locator (VELO) is foreseen to be replaced 
by a silicon hybrid pixel detector \cite{TDR} with an ASIC dubbed ``VeloPix'' 
which will be derived from the Medipix family of ASICs. 
Like its predecessors Medipix2, Timepix, Medipix3, and Timepix3,
the VeloPix chip will 
feature a matrix of $256\times256$ square pixels with a pitch of 55\,\textmu{m}.
Prior to the arrival of the Timepix3 ASIC, 
Timepix and Medipix3 were the most suitable devices for 
the qualification of prototype sensors for the VELO upgrade. 
Until the end of lifetime of the upgraded experiment,
the pixels closest to the beam line
($r=5.1$\,mm) accumulate a fluence of up to
$8\times10^{15}$\,1\,MeV\,n$_{\text{eq}}$ cm$^{-2}$. 
Qualifying silicon sensors in terms of radiation hardness   
is therefore a key element of the VELO upgrade R\&D programme. 

Timepix silicon detectors have been characterised extensively 
in terms of charged-particle tracking performance \cite{Akiba2012}.
Since the ASIC has per-pixel information on the 
collected charge in terms of a time-over-threshold (ToT) value,
a direct measurement of the charge deposition spectrum is possible.

Medipix3 \cite{Llopart2011,Ballabriga2011} on the other hand is a pure counting chip
which was designed primarily for photon imaging applications. 
In order to measure the deposited charge in the sensor one therefore has 
to resort to indirect methods (Section~\ref{Sec:Measurements}). 
Among the large-scale ASICs in the 
high-energy physics community, 
Medipix3 has been the first to be based on IBM 130\,nm CMOS technology. 
Compared to the Timepix (which was fabricated in 250\,nm technology),
it is expected to be more radiation tolerant \cite{Plackett2009} 
and thus lends itself for testing irradiated sensors.

In this work, we report on measurements with 
irradiated and non-irradiated
Medipix3 assemblies carried out in 2012 at the H8 beamline of the 
CERN North Area facility, using positively charged hadrons with 
a momentum of 180~GeV/$c$.
These measurements are intended to provide a validation of 
the chip functionality and performance 
complementary to characterisation measurements using photon sources.
In addition, they also 
represent a first step towards a comprehensive evaluation 
of the radiation hardness of silicon pixel sensors 
with the ``Medipix footprint'' of $55\times55$~\textmu{m}$^{2}$ pixels.

%% file: setup.tex
\section{Setup}\label{Sec:Setup}
\begin{figure}
  \centering
  \includegraphics[width=0.48\textwidth]{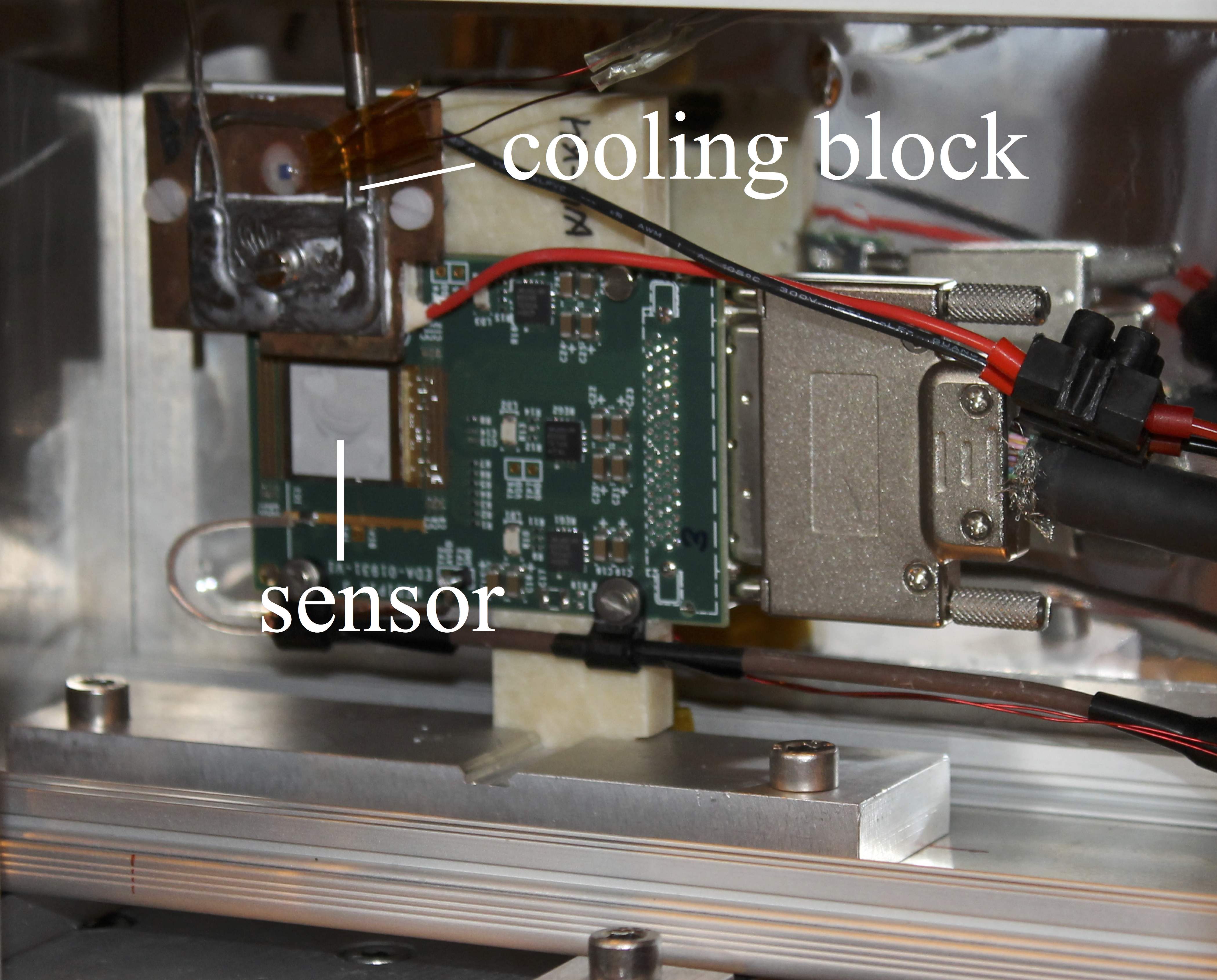}
  \caption{DUT mount and cooling setup. A Peltier cooler (not visible 
           in the photograph) is clamped between the cooling block 
           and the TPG sheet onto which the chip is glued.}
  \label{Fig:Cooling}
\end{figure}
The Timepix telescope, described in Ref.~\cite{Akiba2013}, 
was used for reconstructing the tracks of 
particles crossing the Medipix3 device under test (DUT).
In order to minimise the pointing error, 
the DUT was placed in the centre of the telescope.
For reading out the Medipix3 chip, the  ``Merlin'' data acquisition system 
\cite{Plackett2013} developed at the Diamond Light Source facility was used.
The synchronisation of Timepix and Medipix3 is straightforward, 
as both ASICs work in a ``camera-style'' frame-based readout mode.
An external circuit implemented using NIM modules 
was used for sending shutter opening and closing signals 
to the telescope and the device under test. 
The coincident firing of two scintillators located upstream and downstream 
of the telescope was used for counting the number of beam particles 
traversing the telescope 
and the duration of a shutter was adjusted such that  
50 scintillator triggers were accumulated in one frame. 
During one spill (9.6\,s) typically about 500 frames were recorded.

The temperature of the irradiated samples was controlled 
using a combination of thermoelectric and CO$_{2}$ cooling, 
the latter being provided by a portable cooling plant 
\cite{Verlaat2012}.
As can be seen on the photograph in figure~\ref{Fig:Cooling}, 
the chip was glued onto a tape of Thermal Pyrolytic Graphite (TPG) 
the other end of which was attached to the cold side of 
a Peltier cooler. The hot side of the Peltier cooler was 
put in contact with an aluminium cooling block through which 
CO$_{2}$ was circulating. 
With the chip switched off, a temperature of approximately $-20^{\circ}$\,C 
was reached, rising to $-15^{\circ}$\,C with the chip in operation.
The setup was placed in a light-tight aluminium case which was 
flushed with nitrogen.
The non-irradiated assemblies were measured at room temperature.

In all measurements discussed below, the DUTs were $n$-on-$p$ silicon 
sensors bump-bonded to Medipix3.1 ASICs. 
The ASICs were operated 
in single-pixel high-gain mode and only one threshold (DAC THL) was used.
Before taking data, a threshold equalisation was performed using 
the front-end noise as reference.
The equalisation procedure consists of optimising the DAC values of two global 
current sources (ThresholdN and DACpixel) 
and the threshold adjustment bits of each pixel, 
such that the THL value corresponding to the noise floor lies within a 
certain window for all pixels. 
For the non-irradiated assemblies a target window of $5 < \text{THL} < 25$ 
was used.

Prior to the beam test, calibration measurements using testpulses 
were performed to determine the relation between THL DAC and injected charge.
To verify the viability of the testpulse method,
data were taken with a $^{241}$Am source for the two non-irradiated 
assemblies discussed below, and -- after applying the calibrations obtained 
from the testpulse scan -- the signal peaks were found to match within
2\% between the two assemblies.
These measurements were however made 
with a different readout system \cite{Pixelman} 
and equalisation mask than used in the testbeam (at that time the testpulse 
functionality was not yet implemented in Merlin).
 
Where possible\footnote{One of the assemblies (W20\_B6) was accidentally 
damaged after the beam test.},
calibration measurements using test pulses were later (after the beam test) 
made also using the Merlin system. These calibration curves are used 
in the following for a relative comparison of the signals measured 
with different sensors.

%% file: measurements.tex
\section{Measurements}\label{Sec:Measurements}
For each track reconstructed in the telescope, the intercept 
with the DUT plane is calculated.
In order to suppress fake tracks a requirement on the track 
quality is applied 
and the tracks are required to include hits on all telescope planes.
If an unused cluster with a centre of gravity within a radius 
$r_{w}=110$\,\textmu{m} around 
the track intercept is found, it is associated to the track 
and tagged as used. 
In case of multiple candidate clusters, the closest one is selected.
The hit efficiency (or cluster finding efficiency) 
$\varepsilon$ is then given by the 
fraction of tracks with an associated cluster on the DUT. 
The pointing resolution of the telescope 
($\sim1.5$\,\textmu{m} in the present configuration) 
allows one to probe the hit efficiency as function of the track intercept 
within a pixel cell. In the analysis we divide the pixel cell 
in $9\times9$ bins and calculate separate efficiencies for each bin. 


The most probable value (MPV) of the deposited charge 
can be estimated 
by scanning the hit efficiency as function of threshold. 
Assuming that the distribution of the collected charge can be described 
by a Landau distribution $f_{L}$ convoluted with a Gaussian distribution $f_{G}$, 
the hit efficiency as a function of the threshold $Q$ is given by
\begin{equation}\label{Eq:IntegratedLanGau}
  \varepsilon\left(Q\right) = \int\limits_{Q}^{\infty}\text{d}x f_{L}\otimes f_{G}\left(x\right).
\end{equation}
By fitting the measured efficiency with Eq.~\eqref{Eq:IntegratedLanGau}, 
the MPV and width of the Landau 
distribution and the $\sigma$ of the Gaussian can be determined 
(the mean of the Gaussian is fixed to zero).
This method requires that the entire charge deposited by a track 
is collected by a single pixel. 
We therefore use the hit efficiency in the central 
bin of the $9\times9$ matrix for this measurement.

To determine the spatial resolution, 
the distributions of the residuals between the $x,y$ 
coordinates of the track intercepts and the associated clusters 
are calculated.
The standard deviation of the residual distribution is used as a 
resolution metric
(the pointing error of the telescope
represents only a small correction when subtracted in quadrature).

%% file: results.tex
\section{Results}\label{Sec:Results}

\subsection{Non-irradiated assemblies}\label{Sec:ResultsNonIrradiated}
\begin{figure}[h]
  \centering
  \includegraphics[width=0.48\textwidth]{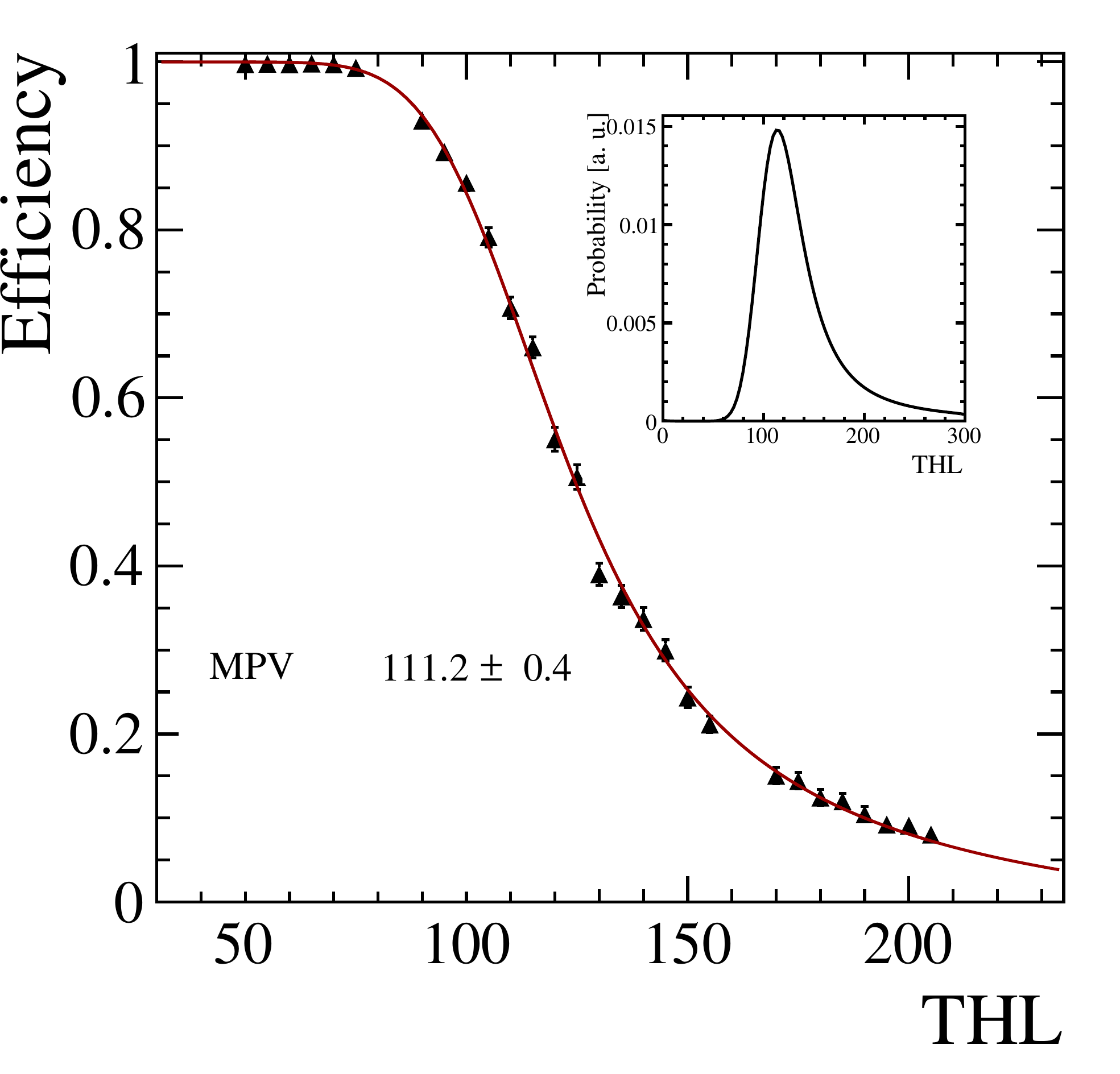}
  \includegraphics[width=0.48\textwidth]{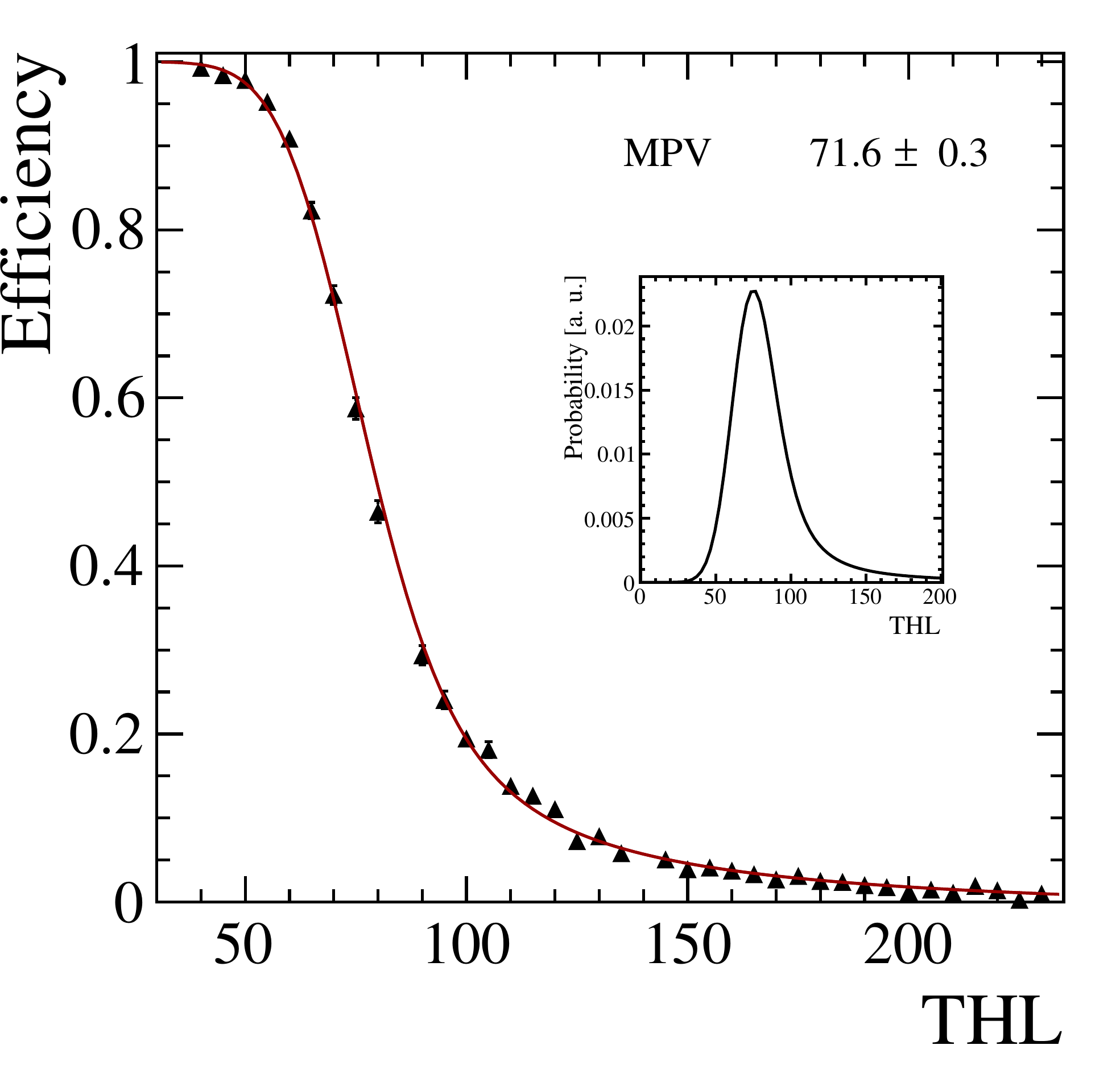}
  \caption{Hit efficiency (for tracks crossing the centre of a pixel cell) 
           in non-irradiated Medipix3 assemblies with 
           100\,\textmu{m} thick $n$-on-$p$ sensors as function of threshold,
           for (left) assembly W20\_B6 and (right) assembly W20\_J9.
           The insets show the distribution of the charge deposition in units of THL.
           The error bars represent the statistical
           uncertainty of the measurements.
           For several points the error bars are smaller than the symbols.}
  \label{Fig:EfficiencyVsThresholdB6J9}
\end{figure}
The measurements before irradiation were 
carried out using $n$-on-$p$ active-edge sensors 
with a nominal thickness of 100\,\textmu{m} 
manufactured by VTT\footnote{VTT Technical Research Centre of Finland, Espoo, Finland}. The MPV of the charge deposition spectrum is thus expected to be 
around $7000 - 7500$\,electrons. 
 
Two assemblies, W20\_B6 and W20\_J9, were tested with beam. 
After the beam test, part of the backside metallisation was 
removed from the sensor on W20\_B6, and 
by injecting laser pulses the depletion voltage $V_{\text{dep}}$
was determined to be approximately $-15$\,V.
In the beam test, both sensors were operated at a bias voltage of $-60$\,V, 
and were oriented perpendicularly to the beam. 
For each point in the threshold scan, a data set comprising 
typically $1-2\times10^{5}$ reconstructed telescope tracks was recorded. 

From the hit efficiency in the centre of the pixel cell 
as function of threshold 
(figure~\ref{Fig:EfficiencyVsThresholdB6J9}), the most probable value of 
the charge deposition spectrum is determined to correspond to 
THL~$\sim111.2$ for assembly W20\_B6 and
THL~$\sim71.6$ for assembly W20\_J9. 
The statistical errors of the fit values are given in figure~\ref{Fig:EfficiencyVsThresholdB6J9}.
Uncertainties due to tracking cuts, alignment, non-linearity of the THL DAC 
give rise to a systematic error on the MPV of $\sim\pm2.5\%$. 
The MPVs in terms of THL DAC values differ significantly between 
the two devices. This can be attributed to non-optimised settings of the 
DAC ``Vcas'' (for voltage cascode), which sets the reference voltage for some transistors in the pixel circuit and, if not tuned properly, can have an impact on the 
operating point of the circuit.

As discussed in section~\ref{Sec:Setup}, testpulse calibration 
measurements using the Merlin readout system could only be made for assembly 
W20\_J9. To facilitate the comparison between the devices, 
the results discussed below are therefore presented as function of the 
threshold-to-signal ratio, i.~e. the applied THL DAC
normalised to the respective MPV. 

\begin{figure}[t]
  \centering
  \includegraphics[width=0.48\textwidth]{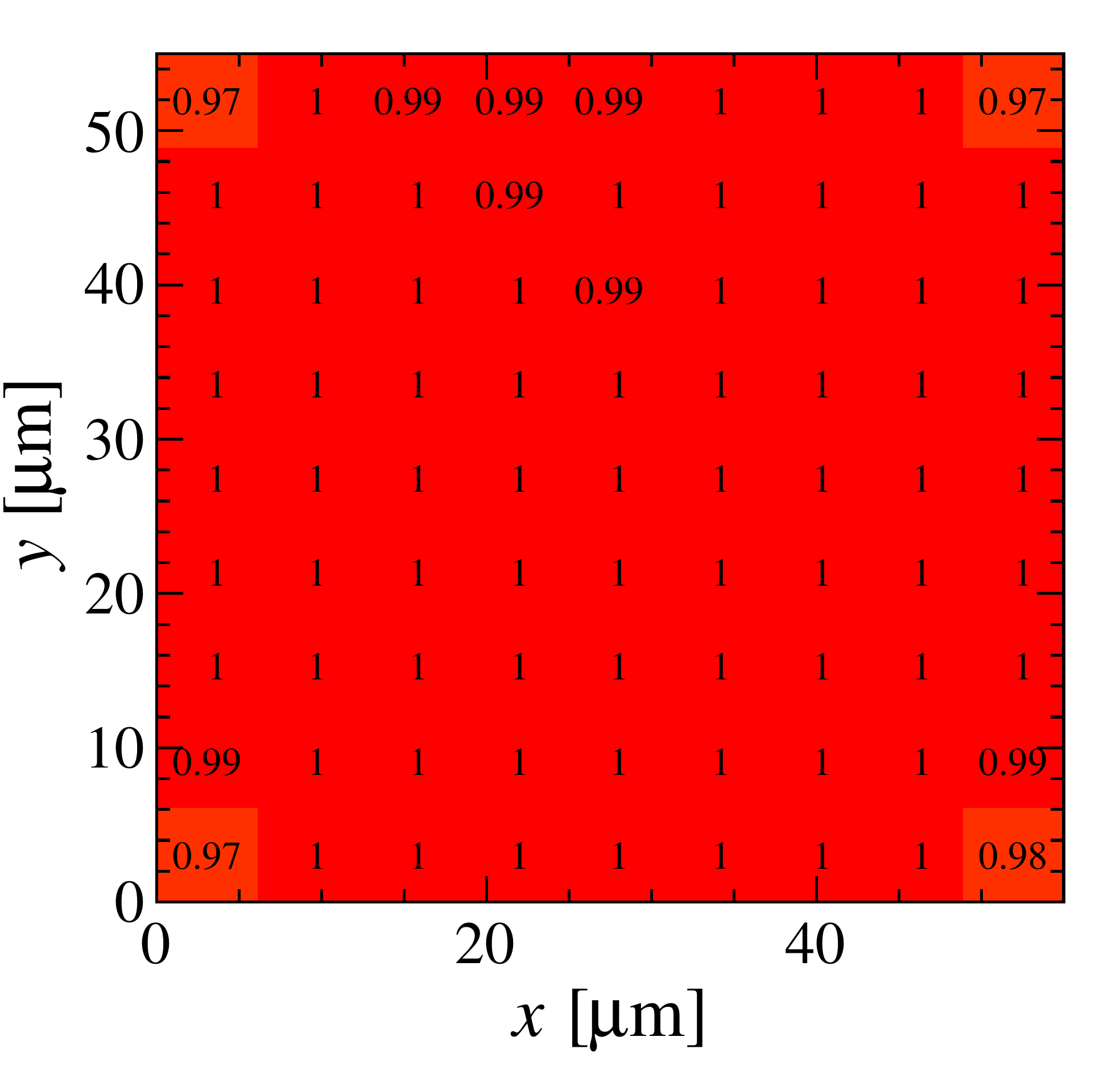}
  \includegraphics[width=0.48\textwidth]{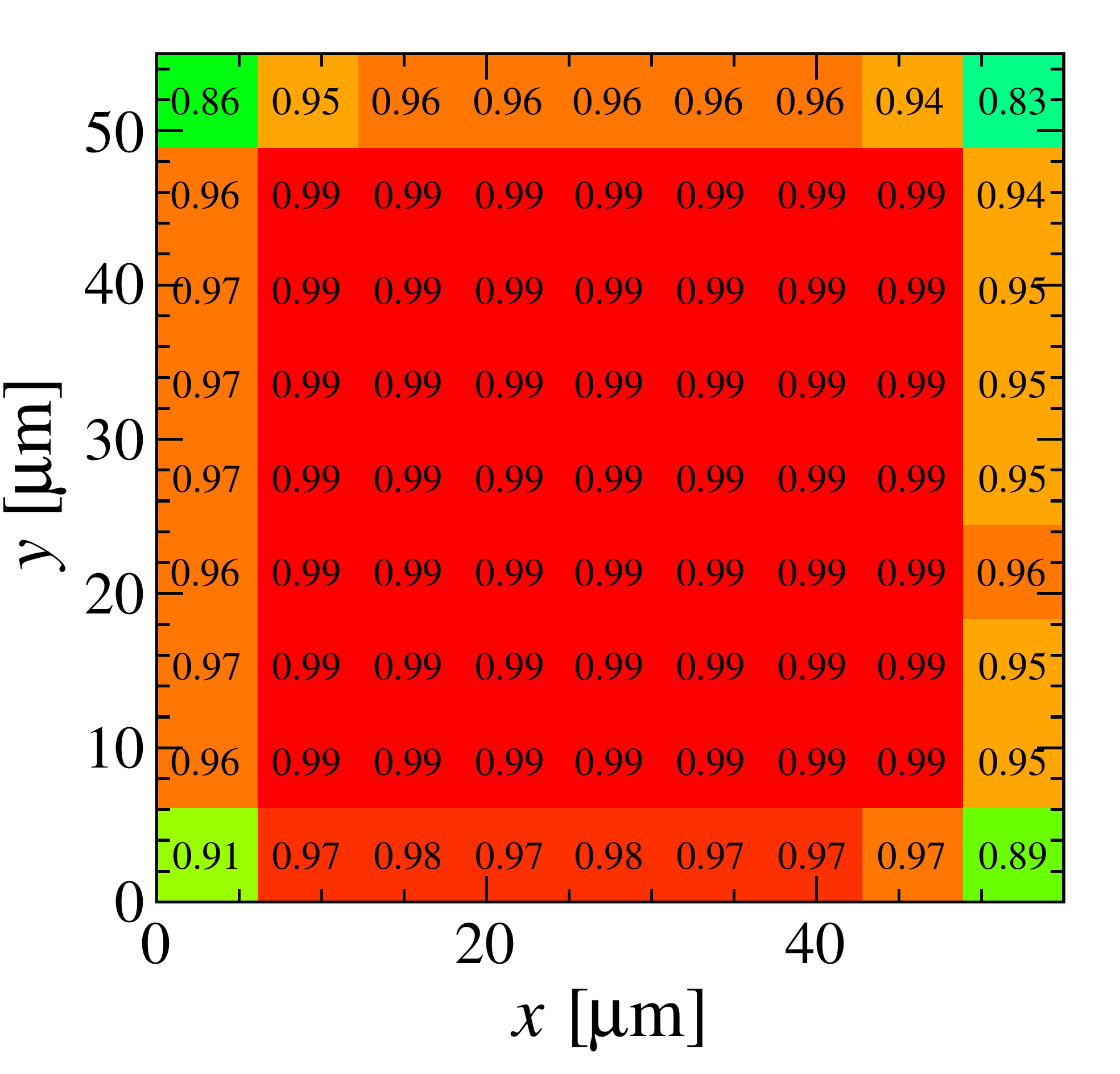}
  \caption{Hit efficiency in non-irradiated Medipix3 assemblies with 
           100\,\textmu{m} thick sensors as function of the track intercept 
	   within a pixel cell, for (left) assembly W20\_B6 at THL/MPV$\sim0.45$ 
           and (right) assembly W20\_J9 at THL/MPV$\sim0.56$.} 
  \label{Fig:EfficiencyVsInterPixelPositionB6J9} 
\end{figure}
\begin{figure}[h]
  \centering
  \includegraphics[width=0.48\textwidth]{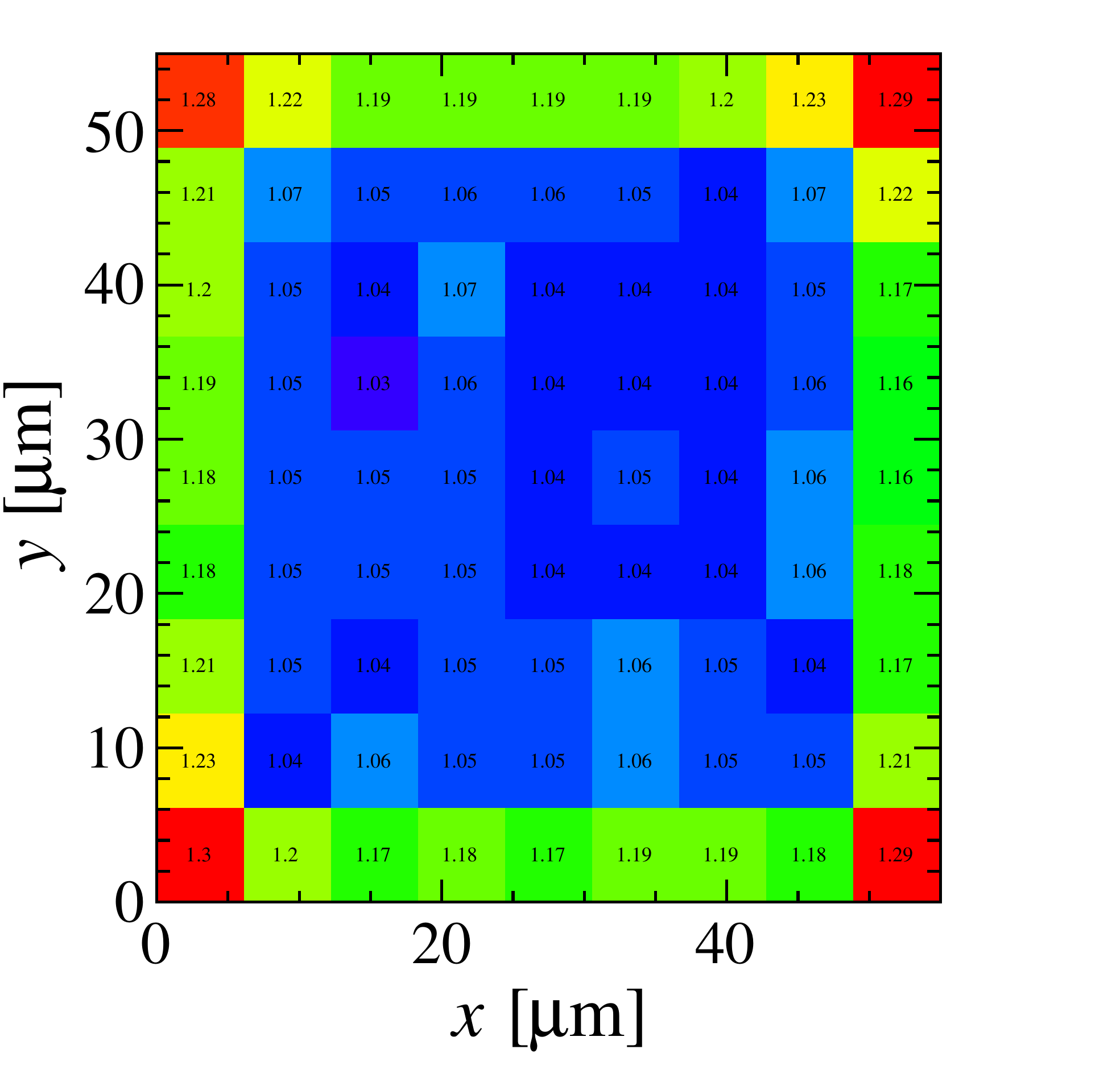}
  \caption{Average cluster size as function of the track intercept within 
           a pixel cell for assembly W20\_B6 (100\,\textmu{m} thick sensor) 
	         at THL/MPV$\sim0.45$. Multi-pixel clusters are found predominantly 
           at the edges and corners of the pixel cell.}
  \label{Fig:ClusterSizeVsInterPixelPosition} 
\end{figure}

Figure~\ref{Fig:EfficiencyVsInterPixelPositionB6J9} (left)  
shows the hit efficiency as function of the track intercept 
within a pixel cell for the lowest threshold-to-signal ratio
(THL/MPV$\sim0.45$) covered by the threshold scan.
At this threshold -- which is high compared to a 
typical operational threshold of 1000\,\,electrons --
the detector can be seen to be fully efficient 
($\varepsilon > 0.99$), except at the corners. 
With increasing threshold --
as illustrated in figure~\ref{Fig:EfficiencyVsInterPixelPositionB6J9} (right)  
-- a drop in efficiency becomes noticeable at the borders of the pixel cell.
This is a consequence of charge sharing due to diffusion, as can be seen from 
figure~\ref{Fig:ClusterSizeVsInterPixelPosition} which shows the average 
cluster size as function of the track intercept within the pixel cell. 

\begin{figure}[h]
  \includegraphics[width=0.48\textwidth]{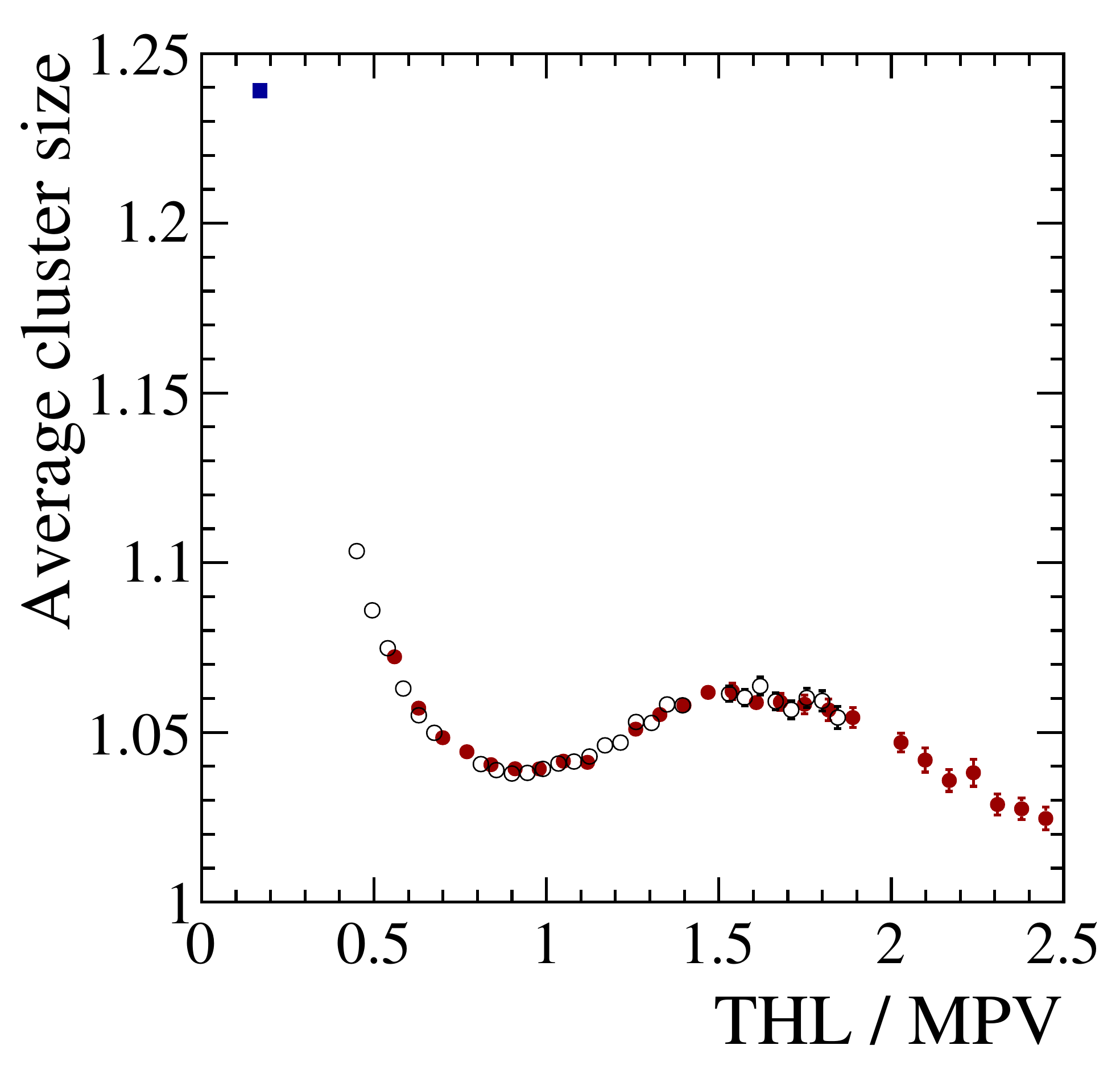}
  \includegraphics[width=0.48\textwidth]{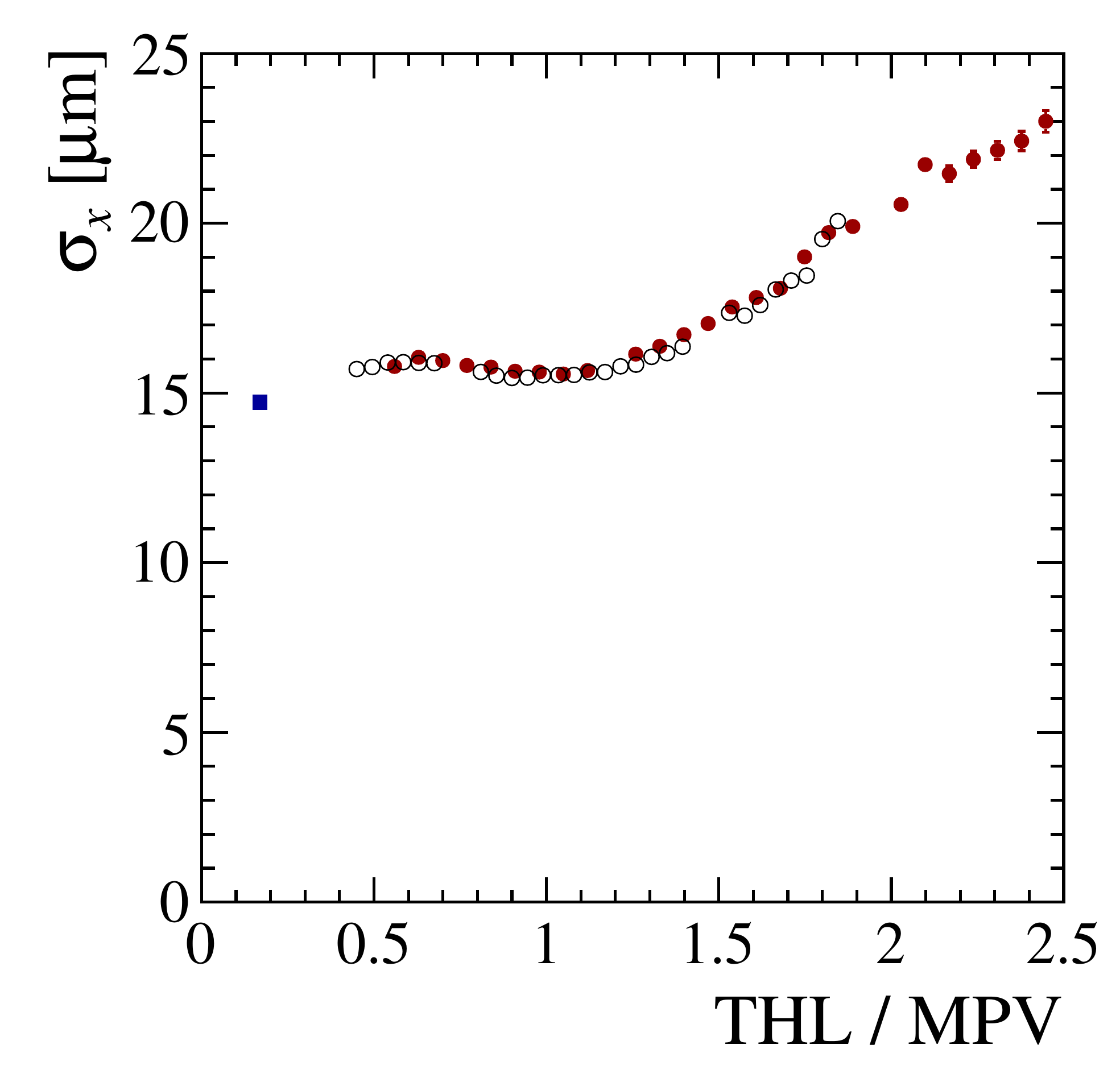}
  \caption{Average cluster size (left) and 
           standard deviation of the $x$-residual distribution (right)
           at perpendicular track incidence 
           as function of threshold-to-signal ratio for 
           assemblies W20\_B6 (empty circles) and W20\_J9 (full circles).
           Results for a Timepix assembly with the same type of sensor are shown with 
           full squares.
           The error bars represent the statistical
           uncertainty of the measurements.
           For most points the error bars are smaller than the symbols.}
  \label{Fig:ClusterSizeResXVsTSJ9B6}
\end{figure}
\begin{figure}[h]
  \centering
  \includegraphics[width=0.48\textwidth]{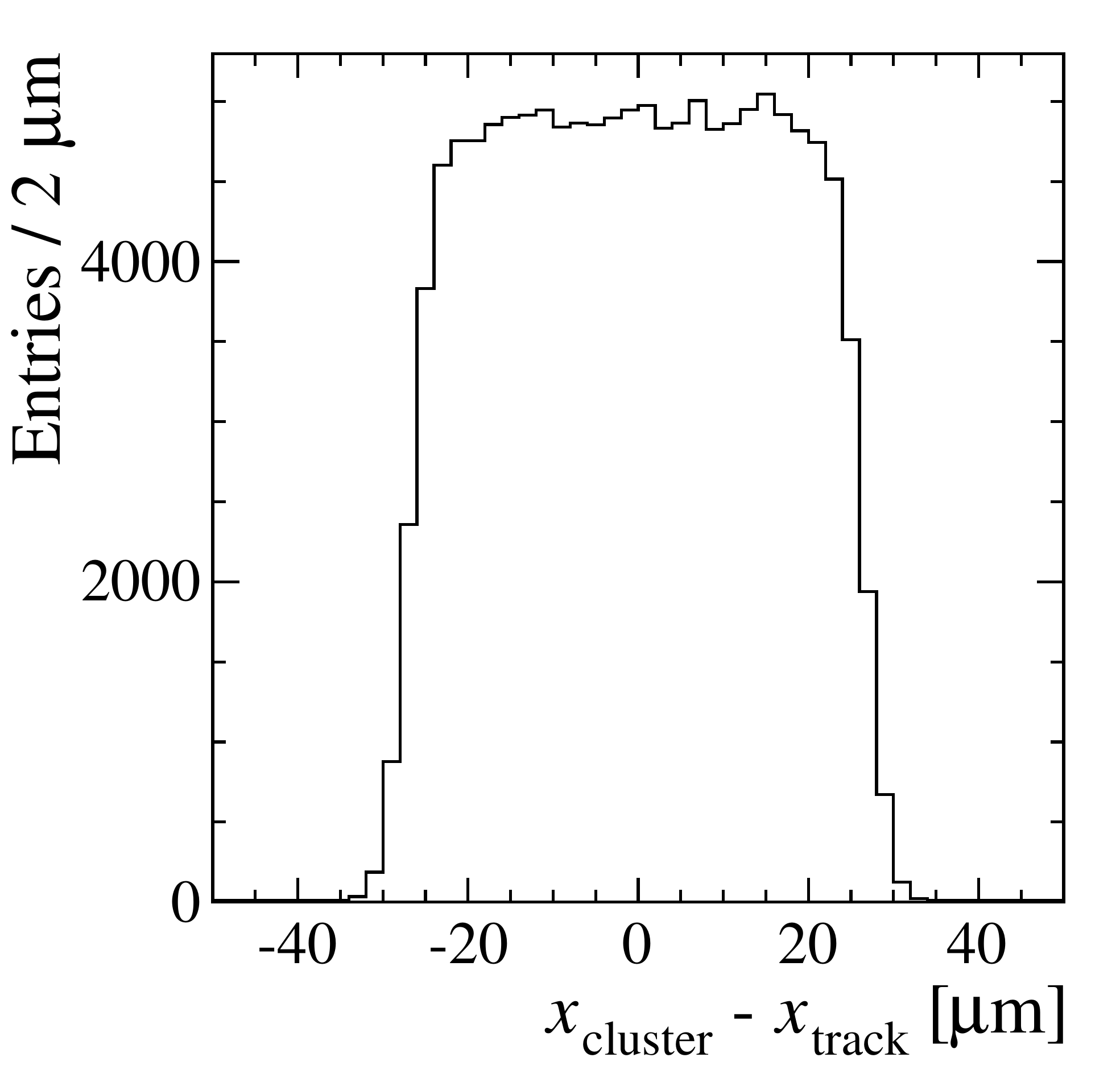}
  \includegraphics[width=0.48\textwidth]{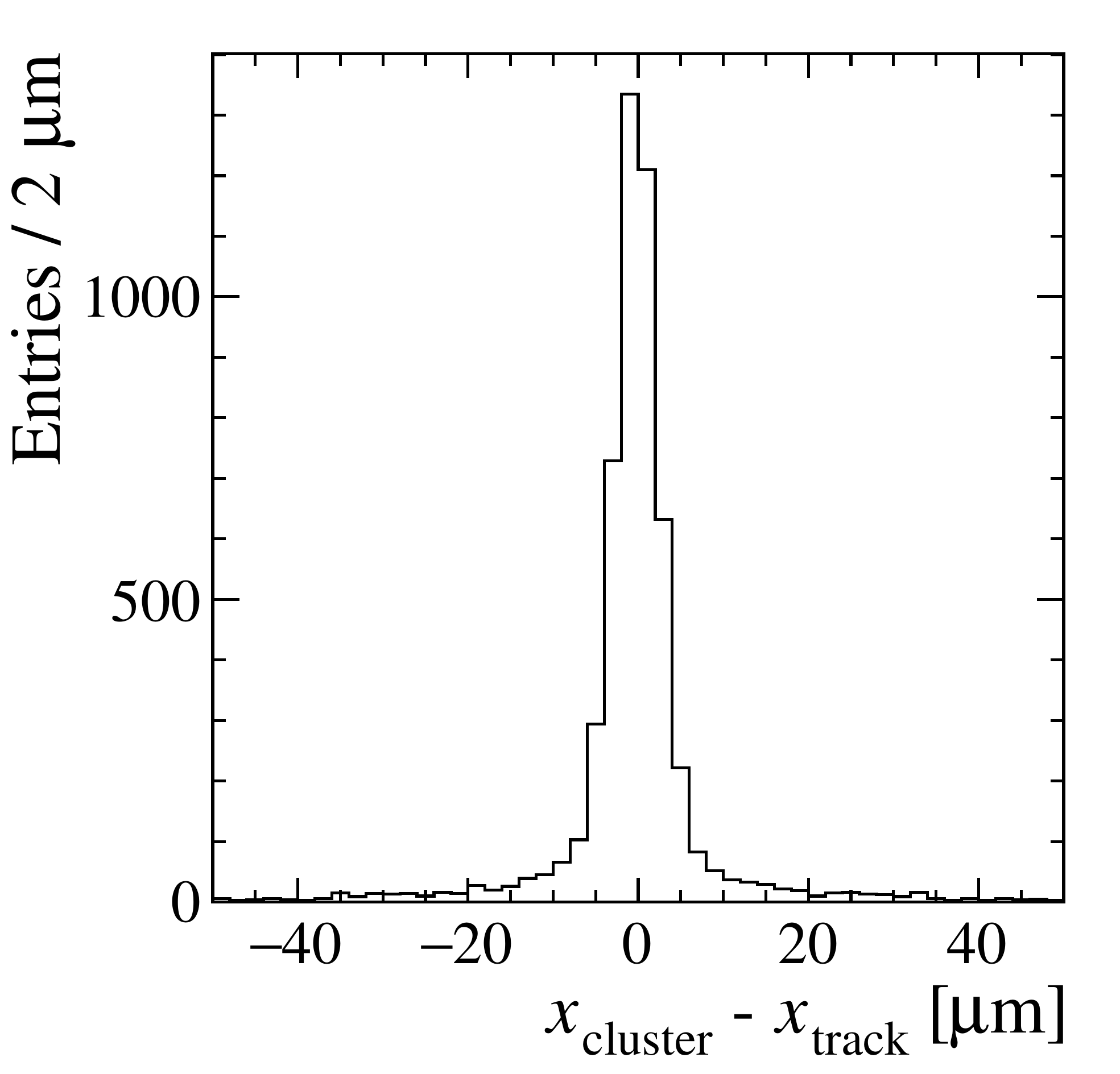}
  \caption{Residual distribution for (left) one-pixel  
           and (right) two-pixel clusters extending over two columns 
           in a non-irradiated 
           100\,\textmu{m} thick sensor (assembly W20\_B6, at THL/MPV$\sim$0.45).}
  \label{Fig:ResidualDistributionJ9}
\end{figure}
Figure~\ref{Fig:ClusterSizeResXVsTSJ9B6} (left) shows the average cluster size 
as function THL/MPV. 
At the lowest threshold, an average cluster size of $1.103\pm0.001$ is found, 
with single-pixel clusters constituting $\sim91.3\%$  
and two-pixel clusters $\sim7.7\%$ of all associated clusters.
The cluster size as function of threshold/signal 
follows the same shape for both sensors, exhibiting a minimum 
around THL/MPV$\sim0.9$ and a subsequent maximum around 
THL/MPV$\sim1.6$.
With increasing threshold, an increasing fraction of the observed clusters 
is produced by primary particles which suffer collisions with large 
energy loss.
These collisions give rise to energetic electrons which further ionise along 
their path and produce electron-hole pairs away from the trajectory 
of the primary particle. 

The residual distributions (in the $x$ direction) at low threshold 
for one-pixel and two-pixel clusters 
are shown in figure~\ref{Fig:ResidualDistributionJ9}. 
Averaged over all cluster sizes, the standard deviation of 
the residual distribution is found to be $\sim15.7$\,\textmu{m} 
in both $x$ and $y$. 
As expected from the dominance of one-pixel clusters, this value is close 
to the binary limit ($55$\,\textmu{m}$/\sqrt{12}\sim15.9$\,\textmu{m}).
Figure~\ref{Fig:ClusterSizeResXVsTSJ9B6} (right) shows the $\sigma$ of the 
residual distribution in $x$ as function of threshold/signal. 
The resolution can be seen to deteriorate significantly 
after the threshold crosses the MPV.

Figure~\ref{Fig:ClusterSizeResXVsTSJ9B6} also includes results obtained 
with a Timepix ASIC bump-bonded to a 100\,\textmu{m} thick $n$-on-$p$ sensor 
from the same batch. The Timepix assembly was measured in the same 
beam test campaign 
and was operated at a threshold of 1000\,electrons 
and a bias voltage of $-60$\,V. 
The data were taken in ToT mode, but the values shown in figure~\ref{Fig:ClusterSizeResXVsTSJ9B6} were calculated with the ToT values set to one to mimic 
the Medipix3 behaviour. 
The results for the Timepix assembly are in agreement with the 
extrapolated results from the Medipix3 assemblies. 
%

\begin{figure}[h]
  \includegraphics[width=0.48\textwidth]{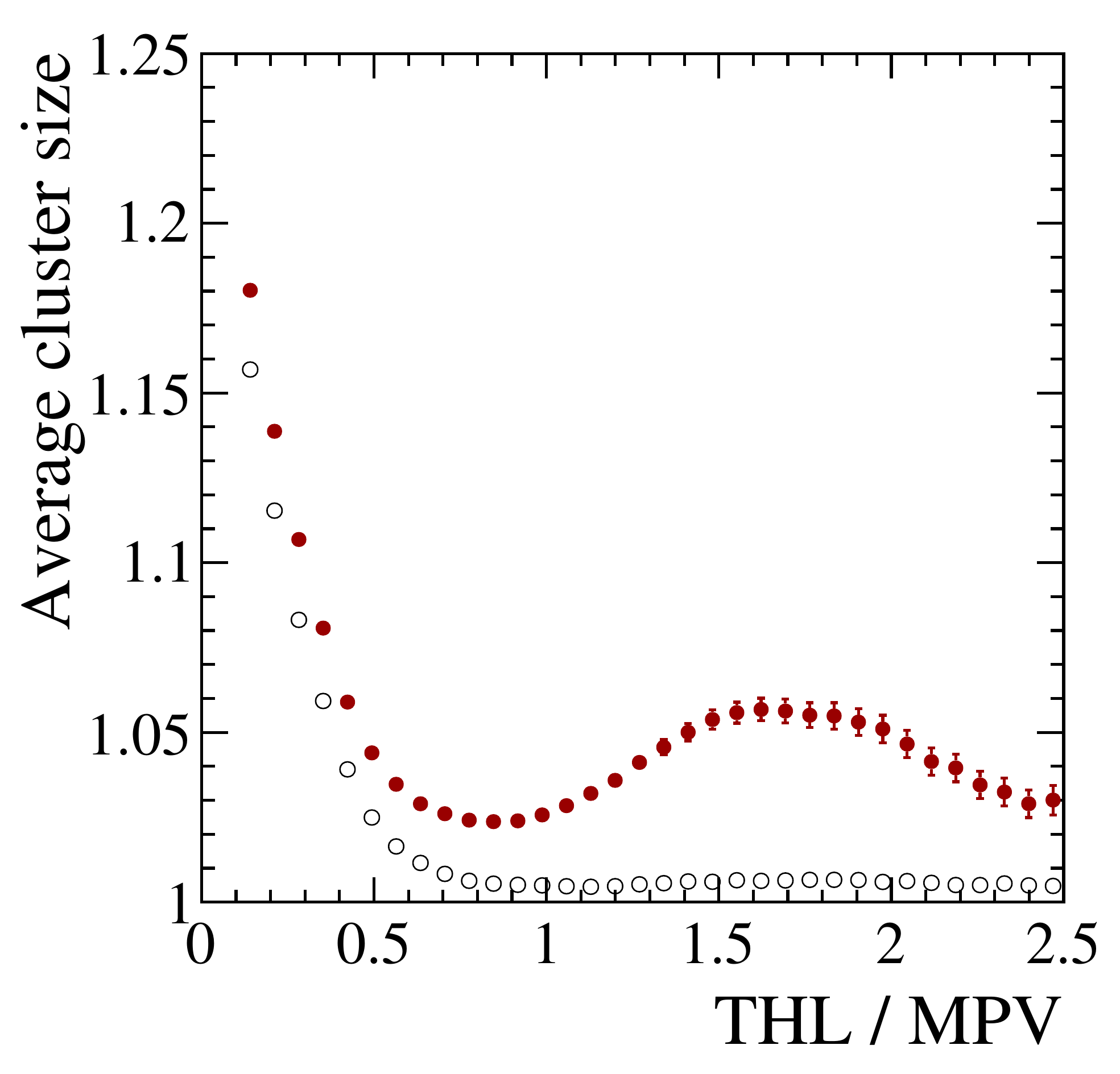}
  \includegraphics[width=0.48\textwidth]{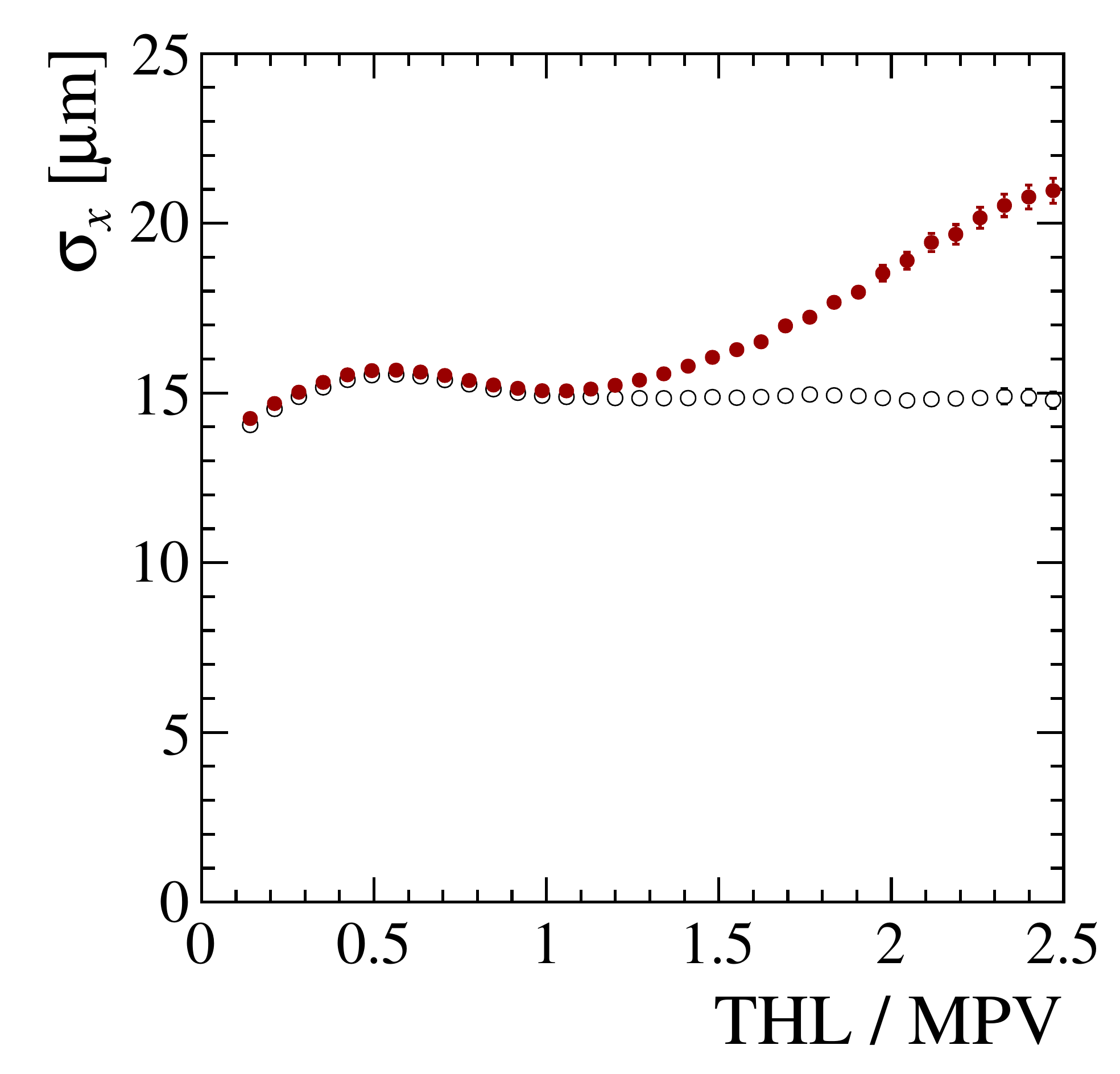}
  \caption{Simulated average cluster size (left) and 
           standard deviation of the $x$-residual distribution (right)
           at perpendicular track incidence 
           as function of threshold-to-signal ratio.
           Full (empty) symbols show the results with (without) the 
           spatial extent of the 
           ionisation pattern being included in the simulation.} 
  \label{Fig:ClusterSizeResXVsTSSim}
\end{figure}
To understand better the observed shapes of resolution and cluster size as 
function of threshold,  
a simple simulation using the Garfield++ toolkit \cite{Garfield} was used. 
The primary ionisation process is calculated 
using the Heed program \cite{Smirnov2005}, 
which in addition to the energy loss by the traversing charged 
particle also simulates the ionisation cascade from high-energy 
(``delta'') electrons produced in the interactions of the 
charged particle with the silicon medium
as well as the spatial distribution of the resulting electron-hole pairs.
Each electron is subsequently transported through the sensor,
based on the drift velocity and diffusion coefficient as function 
of the electric field. A one-dimensional approximation for the 
electric field is used\footnote{The electric field is assumed 
to vary linearly between $E=\left(V - V_{\text{dep}}\right)/d$ 
at the sensor backside and $E=\left(V + V_{\text{dep}}\right)/d$ 
at the implants, where $d$ is the sensor thickness.}. 

As can be seen from figure~\ref{Fig:ClusterSizeResXVsTSSim}, 
the features in the measured cluster size and resolution as 
function of threshold-to-signal ratio are reproduced by the simulation, 
provided that the spatial extent of the ionisation pattern is taken into 
account. This corroborates the conclusion that these features are due 
to ``delta'' electrons.

\subsection{Irradiated assemblies}
\begin{figure}[h]
  \centering
  \includegraphics[width=0.48\textwidth]{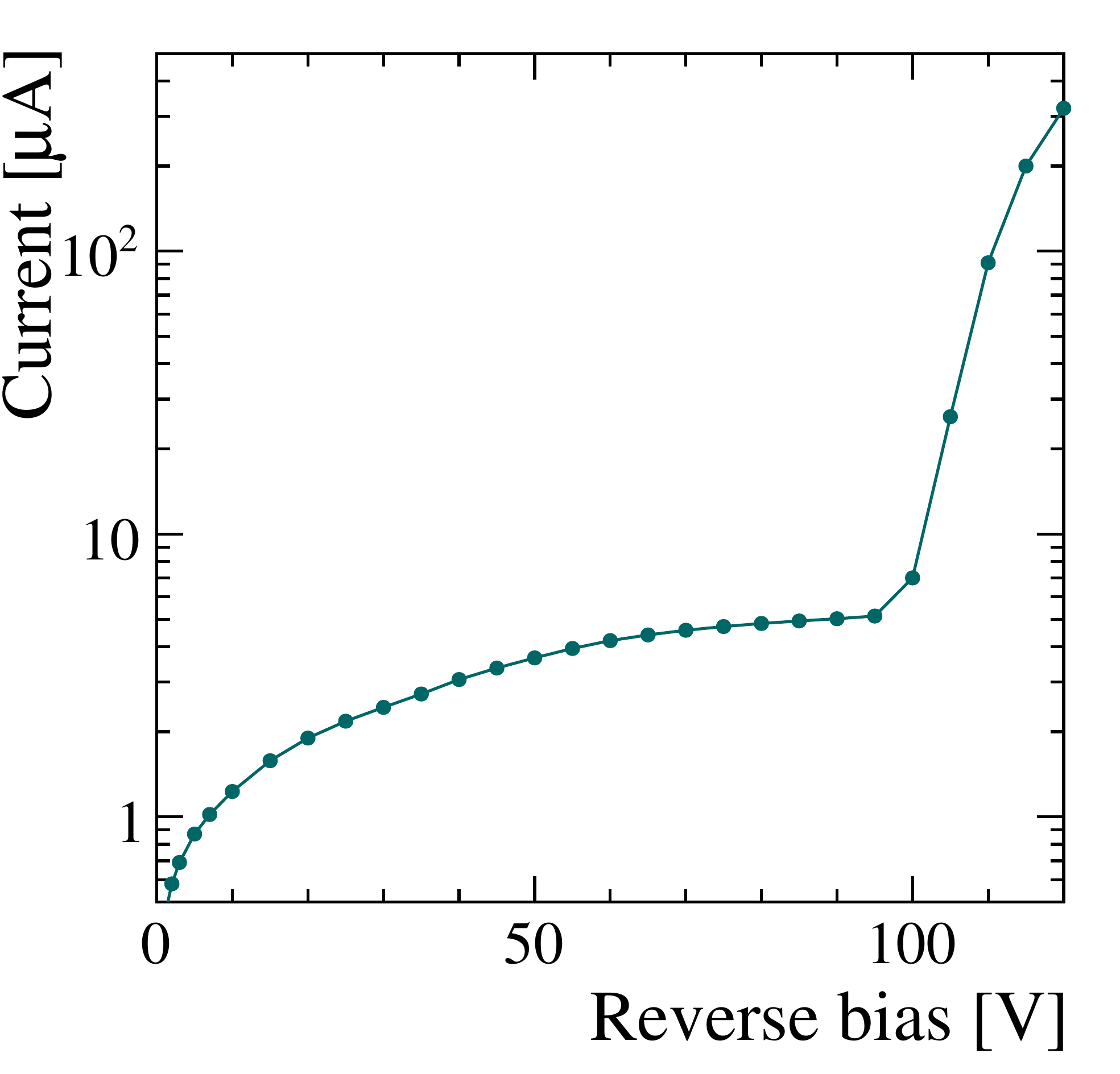}
  \caption{Leakage current as function of reverse bias voltage 
  at $T=-13^{\circ}$\,C after irradiation to $0.5\times10^{15}$\,1\,MeV\,n$_{\text{eq}}$ cm$^{-2}$
  (100\,\textmu{m} thick $n$-on-$p$ active-edge sensor with 100\,\textmu{m} pixel-to-edge distance).}
  \label{Fig:CurrentVsVoltageH5}
\end{figure}
\begin{figure}[h]
  \centering
  \includegraphics[width=0.48\textwidth]{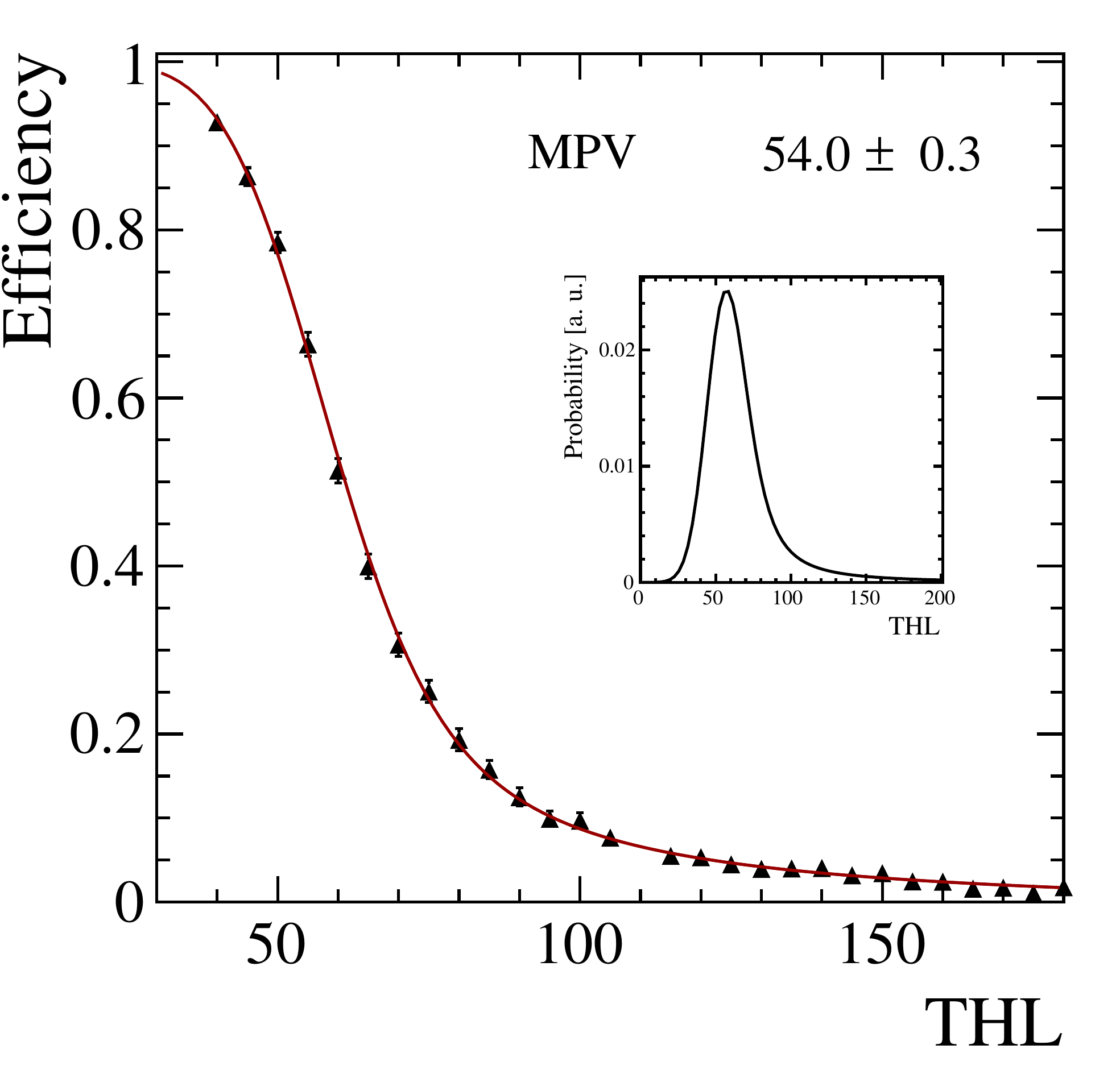}
  \caption{Hit efficiency in Medipix3 assembly W20\_H5 with 
           100\,\textmu{m} thick $n$-on-$p$ sensor as function of threshold, 
            after irradiation to 
            $0.5\times10^{15}$\,1\,MeV\,n$_{\text{eq}}$ cm$^{-2}$.
            The inset shows the distribution of the charge deposition in units of THL 
            (Landau distribution convoluted with a Gaussian).}
  \label{Fig:EfficiencyVsThresholdH5}
\end{figure}
\begin{figure}[h]
  \centering
  \includegraphics[width=0.48\textwidth]{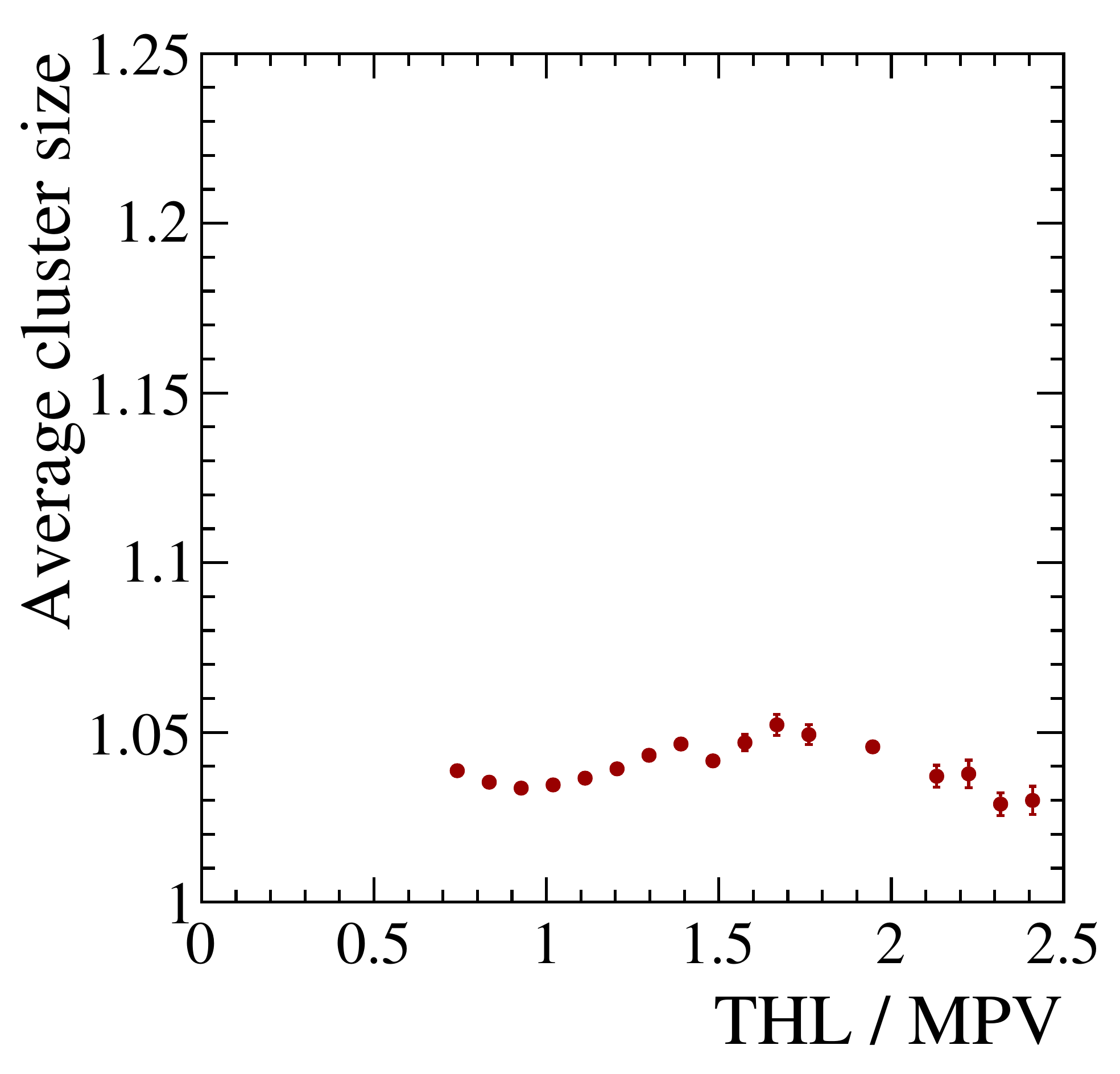}
  \includegraphics[width=0.48\textwidth]{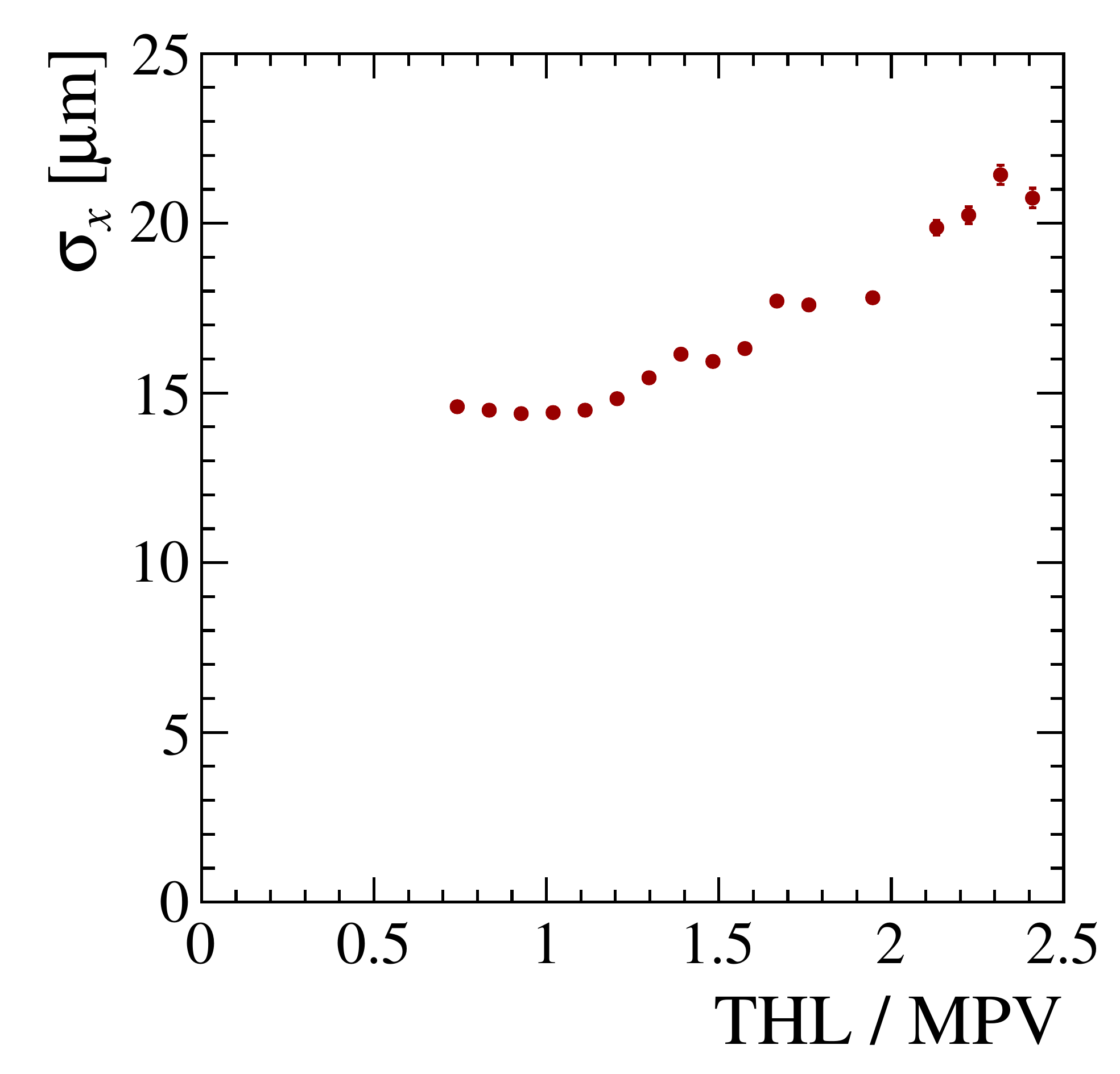}
  \caption{Average cluster size (left) and
           standard deviation of the $x$-residual distribution (right)
           at perpendicular track incidence 
           as function of threshold-to-signal ratio for irradiated
           assembly W20\_H5.}
  \label{Fig:ClusterSizeResXvsThresholdH5}
\end{figure}
A set of Medipix3.1 assemblies, with VTT $n$-on-$p$ active-edge sensors 
from the same batch 
as the non-irradiated sensors discussed above,
were irradiated at the Ljubljana TRIGA reactor \cite{Snoj2012}
to a 1\,MeV neutron equivalent fluence of $5\times10^{14}$~cm$^{-2}$ 
and subsequently annealed for 80 minutes at $60^{\circ}$~C.
From lab measurements with a laser,
the effective depletion voltage $V_{\text{dep}}$ of these sensors 
after irradiation was measured to be between $-85$ and $-105$\,V.
One of the assemblies, W20\_H5, was characterised in the beam test.  
The sensor, which has a pixel-to-edge distance of 100\,\textmu{m}, 
was operated at a bias voltage of $-100$\,V as 
operation at higher bias was inhibited by the onset of electrical breakdown 
(figure~\ref{Fig:CurrentVsVoltageH5}). 

In general, the irradiated assemblies exhibited a higher dark count rate 
compared to the non-irradiated ones, 
which was attributed to electrons from the $\beta$-decay of $^{182}$Ta 
produced by neutron activation during the irradiation of the ASIC\footnote{
The presence of $^{182}$Ta in the irradiated assemblies was confirmed by 
gamma spectroscopy measurements performed by the CERN radioprotection group.}.
This -- presumably in combination with other radiation effects -- 
resulted in problems during the equalisation 
procedure for some assemblies, as well as a larger threshold dispersion. 
The optimised values of the DACs which control the global currents 
(ThresholdN and DACpixel) were larger (by a factor 2 -- 3) 
compared to the non-irradiated assemblies. 

As can be seen from figure~\ref{Fig:EfficiencyVsThresholdH5}, 
the MPV of the charge deposition spectrum corresponds 
to a THL DAC of $\sim53.9$. After correcting 
for the differences in gain and offset using testpulse calibration curves, 
this value is found to be approximately 8.5\% lower than the MPV 
of the non-irradiated sensors.

As in the non-irradiated case, the cluster size spectrum is dominated 
by single-pixel clusters. Cluster size and $\sigma$ of the $x$-residual 
distribution as function of threshold-to-signal ratio 
(figure~\ref{Fig:ClusterSizeResXvsThresholdH5}) follow closely 
the corresponding curves for the non-irradiated assemblies
(figure~\ref{Fig:ClusterSizeResXVsTSJ9B6}).

\begin{figure}[h]
  \centering
  \includegraphics[width=0.48\textwidth]{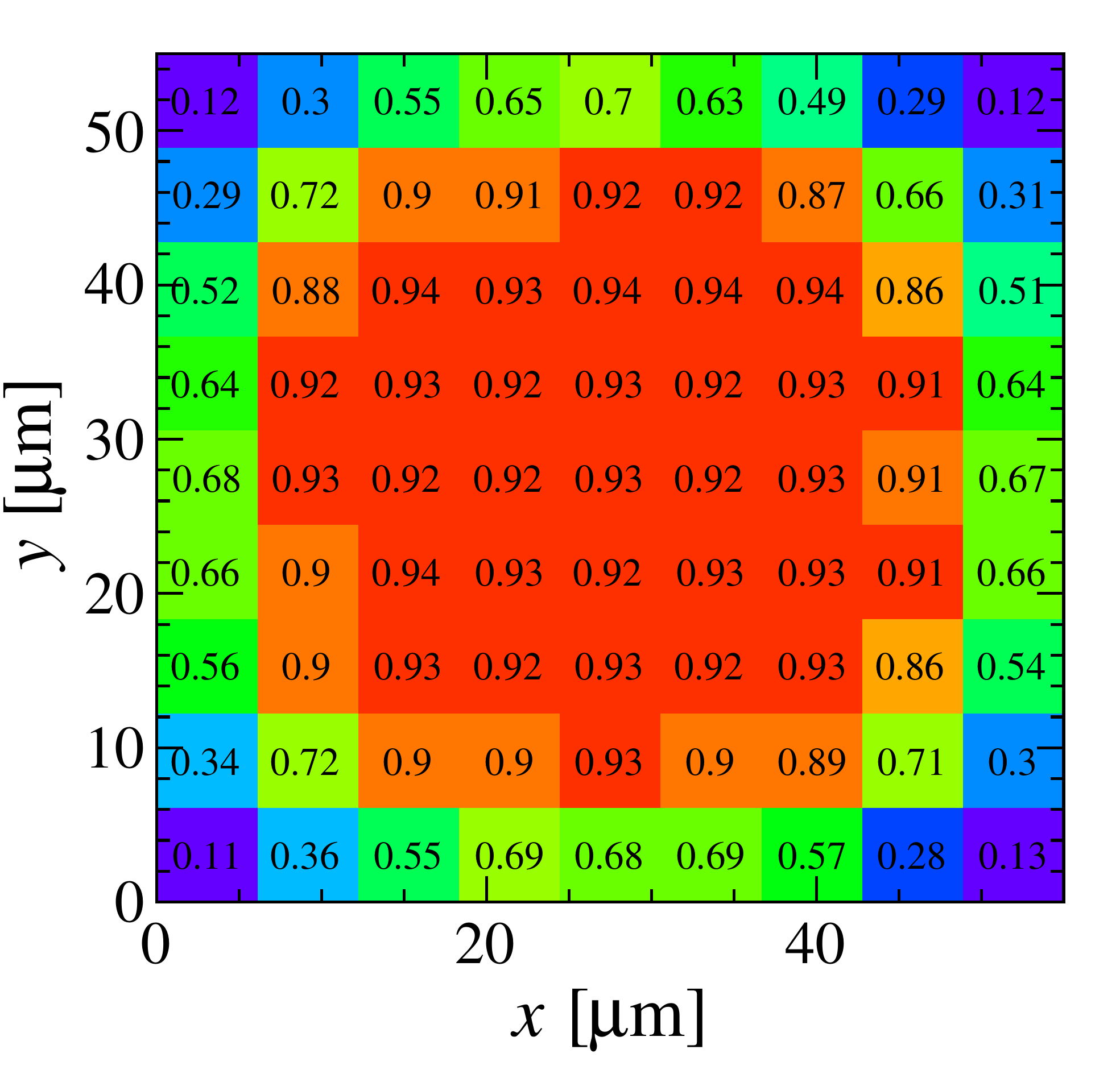}
  \caption{Hit efficiency in Medipix3 assembly W20\_H5 with 
           100\,\textmu{m} thick $n$-on-$p$ sensor as function of the track intercept 
	         within a pixel cell at THL = 40 (threshold/signal\,$\sim0.74$),
           after irradiation to
           $0.5\times10^{15}$\,1\,MeV\,n$_{\text{eq}}$ cm$^{-2}$.
           The drop in efficiency at the edges of the pixel cell is 
           more pronounced than in 
           figure~\protect\ref{Fig:EfficiencyVsInterPixelPositionB6J9} mainly because 
           of the higher threshold-to-signal ratio.}
  \label{Fig:EfficiencyVsInterPixelPositionH5} 
\end{figure}
Figure~\ref{Fig:EfficiencyVsInterPixelPositionH5} shows the hit efficiency 
as function of the track intercept within a pixel cell at the 
lowest measured threshold ($\text{THL}=40$). In the centre of the 
pixel cell, an efficiency of $0.93\pm0.01$ is found, compared to 
$0.97\pm0.01$ for the non-irradiated assemblies at the same 
threshold-to-signal ratio ($\sim0.74$).


\begin{figure}[h]
  \centering
  \includegraphics[width=0.48\textwidth]{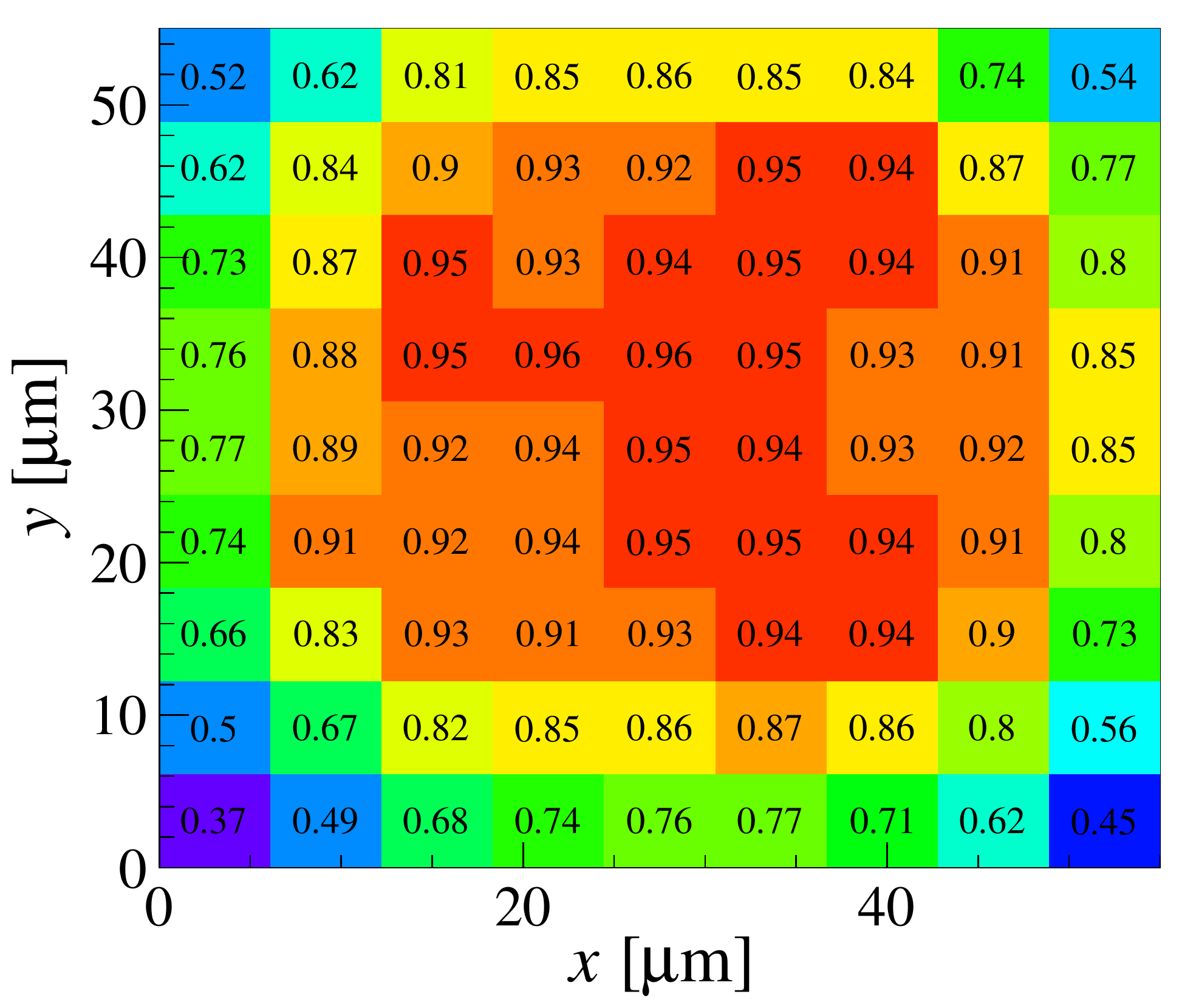}
  \caption{Hit efficiency in Medipix3 assembly W20\_D6 
           with 200\,\textmu{m} thick $n$-on-$p$ sensor 
           (diced at 400\,\textmu{m} from the last pixel) 
           as function of the track intercept within a pixel cell 
           at THL~=~50 and 300\,V bias voltage, 
           after irradiation to 
           $2.5\times10^{15}$\,1\,MeV\,n$_{\text{eq}}$ cm$^{-2}$.
           The asymmetry of the efficiency profile 
           is attributed to a small inclination of the assembly 
           with respect to the beam.}
  \label{Fig:EfficiencyVsInterPixelPositionD6} 
\end{figure}
A further beam test measurement was performed with a Medipix3.1 ASIC 
bump-bonded to a 200\,\textmu{m} thick $n$-on-$p$ sensor manufactured 
by CNM\footnote{Centro Nacional de Microelectr\'onica, Barcelona, Spain}. 
The sensor featured two guard rings and was diced at a distance of 
400\,\textmu{m} from the border of the pixel matrix. 
The assembly (W20\_D6) 
was also irradiated at Ljubljana, but to a higher fluence, 
$2.5\times10^{15}$\,1\,MeV\,n$_{\text{eq}}$ cm$^{-2}$.
During the equalisation procedure, a large threshold dispersion was observed,
such that the upper limit of the equalisation target window needed to 
be increased to $\text{THL}=40$. 
In addition, a significant fraction ($\sim10\%$) 
of the pixels needed to be masked.
Tracks crossing a masked pixel were thus excluded from the efficiency 
measurements for this device. 

Because of time constraints, data were taken only at three THL values 
with this device, such that determining the MPV of the 
charge deposit spectrum was not possible.

\begin{figure}[h]
  \centering
  \includegraphics[width=0.48\textwidth]{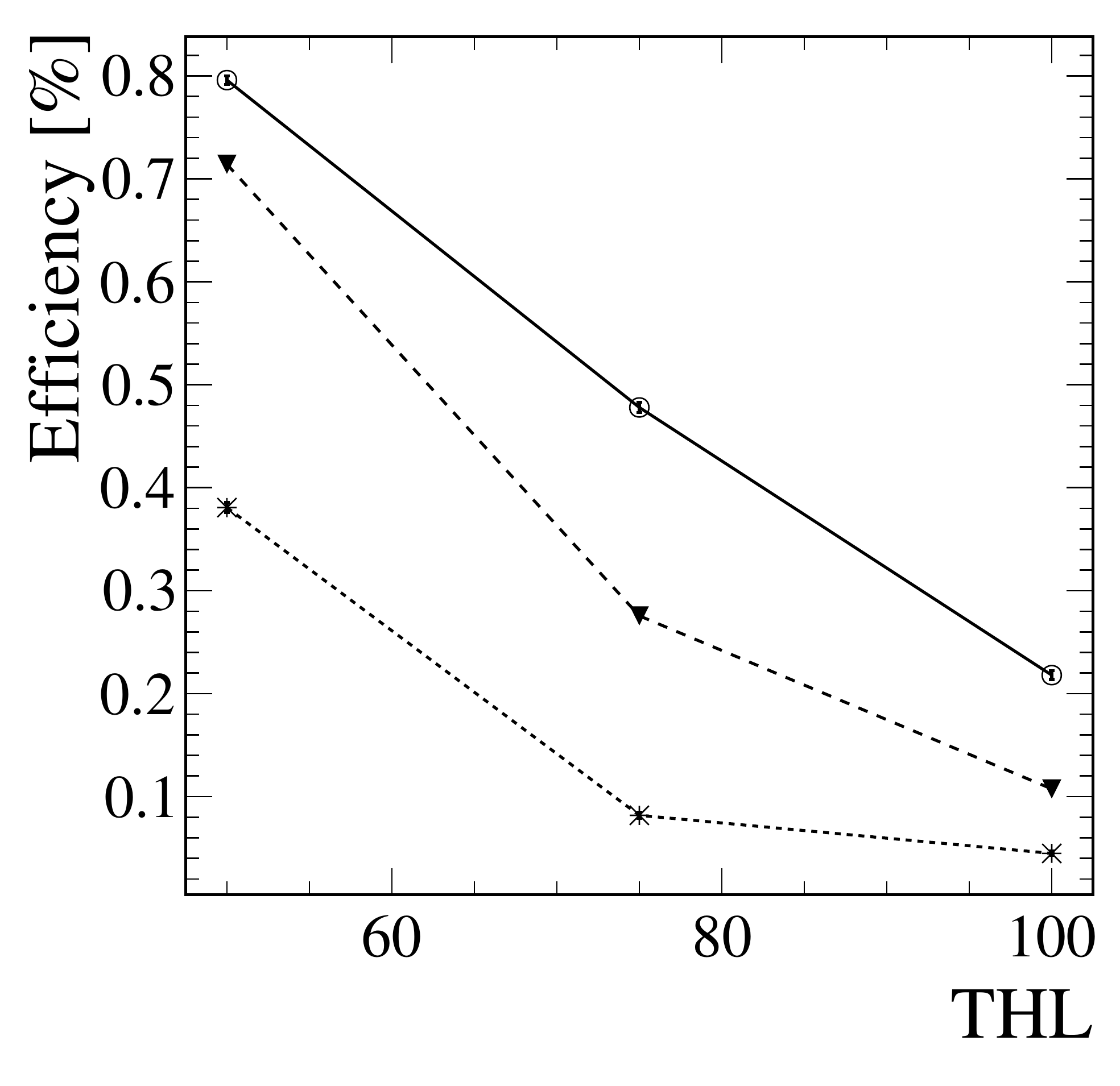}
  \caption{Overall hit efficiency in irradiated assembly W20\_D6 
  with 200\,\textmu{m} thick $n$-on-$p$ sensor 
  as function of the THL value for different bias voltages 
  (stars: $-100$\,V, triangles: $-200$\,V, circles: $-300$\,V). 
  The assembly was irradiated to a fluence of 
  $2.5\times10^{15}$\,1\,MeV\,n$_{\text{eq}}$ cm$^{-2}$. 
  The lines are drawn to guide the eye.}
  \label{Fig:EfficiencyVsBiasVsTHLD6} 
\end{figure}

Figure~\ref{Fig:EfficiencyVsInterPixelPositionD6} shows 
the efficiency as function of the track intercept. 
The dependency of the efficiency on the applied threshold and 
bias voltage is shown in figure~\ref{Fig:EfficiencyVsBiasVsTHLD6}.

\clearpage

%% file: conclusions.tex
\section{Conclusions}
First results on the performance of 
hadron-irradiated pixel detectors with the ``footprint'' of the 
upgraded LHCb VELO ($55\times55$\,\textmu{m}$^{2}$ pixels) 
have been obtained from a beam test at the CERN SPS, 
using Medipix3.1 ASICs bump-bonded to thin $n$-on-$p$ silicon sensors. 
As the Medipix3 ASICs do not provide a direct measurement of the 
collected charge, threshold scans were used for characterising 
the sensor.

Reference measurements with non-irradiated Medipix3.1 assemblies were made in 
the same testbeam campaign. 
The threshold DAC value (THL) corresponding to the 
most probable value of the charge deposition spectrum was determined 
from the hit efficiency as function of threshold. 
The measured cluster size and resolution as function of the 
THL/MPV ratio are consistent between the assemblies and in good agreement 
with simulations. 

After irradiation with reactor neutrons to a fluence of 
$0.5\times10^{15}$\,1\,MeV\,n$_{\text{eq}}$ cm$^{-2}$, 
a larger threshold dispersion and 
an increased power consumption of the ASIC were observed. 
A direct comparison between the performance of irradiated and non-irradiated 
sensors is complicated due to the fact that the non-irradiated sensors 
were operated at overdepletion while the irradiated sensor (due to early 
electrical breakdown) needed to be operated close to depletion. 
The difference in operating conditions may explain the small drop in 
efficiency and MPV measured with the irradiated sensor. 
The shapes of average cluster size and resolution versus threshold 
are not changed significantly compared 
to the non-irradiated sensors.



The studies presented in this paper 
are being continued using the Timepix3 ASIC, 
which has recently become available. 

%% file: acknowledgements.tex
\section*{Acknowledgements}
We gratefully acknowledge strong support from the CERN Medipix Group.
in particular Rafael Ballabriga. 
We would also like to thank Juha Kalliopuska and Sami V\"ah\"anen from 
VTT/Advacam and Giuglio Pellegrini from the IMB-CNM Radiation Detectors Group. 
The research leading to these results has received partial funding from the European Commission under the FP7 Research Infrastructures project AIDA, grant agreement no. 262025. 
We gratefully acknowledge the expert wire bonding support provided by Ian McGill of the CERN DSF bonding lab.
We extend warm thanks to Igor Mandi\'{c} and Vladimir Cindro for irradiating 
the Medipix3 assemblies.

%% file: references.tex

%% file: main.bbl
\begin{thebibliography}{99}
  \bibitem{TDR}
  LHCb collaboration, R. Aaij et al., 
  \emph{LHCb VELO Upgrade Technical Design Report}, 
  CERN-LHCC-2013-021. LHCB-TDR-013.
  \bibitem{Akiba2012}
  K. Akiba et al., 
  \emph{Charged particle tracking with the Timepix ASIC},
  \emph{Nucl. Instr. Meth. A} \textbf{661} (2012), 31--49,
  [\href{http://arxiv.org/abs/1103.2739}{arXiv:1103.2739}]
  \bibitem{Llopart2011}
  R. Ballabriga, M. Campbell, E. Heijne, X. Llopart, L. Tlustos, 
  and W. Wong, 
  \emph{Medipix3: A 64 k pixel detector readout chip working in 
  single photon counting mode with improved spectrometric performance}, 
  \emph{Nucl. Instr. Meth. A} \textbf{633} (2011), S15 -- S18
  \bibitem{Ballabriga2011}
  R. Ballabriga et al., 
  \emph{Characterization of the Medipix3 pixel readout chip},
  \jinst{6}{2011}{C01052}
  \bibitem{Plackett2009}
  R. Plackett, X. Llopart, R. Ballabriga, M. Campbell, L. Tlustos, and W. Wong, 
  \emph{Measurement of Radiation Damage to 130 nm Hybrid Pixel Detector Readout Chips}, 
  Proceedings of Topical Workshop on Electronics for Particle Physics (TWEPP09), CERN-2009-006
  \bibitem{Akiba2013}
  K. Akiba et al.,
  \emph{The Timepix Telescope for high performance particle tracking},
  \emph{Nucl. Instr. Meth. A} \textbf{723} (2013), 47--54,
  [\href{http://arxiv.org/abs/1304.5175}{arXiv:1304.5175}]
  \bibitem{Plackett2013}
  R. Plackett, I. Horswell, E. N. Gimenez, J. Marchal, D. Omar, 
  and N. Tartoni,
  \emph{Merlin: a fast versatile readout system for Medipix3},
  \jinst{8}{2013}{C01038}
  \bibitem{Verlaat2012}
  B. Verlaat, L. Zwalinski, R. Dumps, M. Ostrega, P. Petagna, and T. Szwarc,
  \emph{TRACI, a multipurpose CO$_2$ cooling system for R\&D},
  10$^{\text{th}}$ IIR-Gustav Lorentzen Conference on Natural Working Fluids (2012), Refrigeration Science and Technology Proceedings 2012-1
  \bibitem{Pixelman}
  D. Ture\v{c}ek et al.,
  \emph{Pixelman: a multi-platform data acquisition and processing software package for Medipix2, Timepix and Medipix3 detectors},
  \jinst{6}{2011}{C01046}
  \bibitem{Garfield}
  \emph{Garfield++ -- simulation of tracking detectors}, 
  \href{http://cern.ch/garfieldpp}{http://cern.ch/garfieldpp} 
  \bibitem{Smirnov2005}
  I. B. Smirnov, 
  \emph{Modeling of ionization produced by fast charged particles in gases},
  \emph{Nucl. Instr. Meth. A} \textbf{554} (2005), 474 -- 493
  \bibitem{Snoj2012}
  L. Snoj, G. \v{Z}erovnik, A. Trkov, 
  \emph{Computational analysis of irradiation facilities at the JSI TRIGA reactor}, 
  \emph{Applied Radiation and Isotopes} \textbf{70} (2012), 483--488
\end{thebibliography}
